\UseRawInputEncoding
\documentclass[superscriptaddress, twocolumn, amsmath, amssymb, aps,prl, notitlepage,longbibliography,10pt]{revtex4-1} 
\usepackage{graphicx,graphics,epsfig,subfigure,times,bm,bbm,amssymb,amsmath,amsfonts,amsthm,mathrsfs,MnSymbol,physics}
\usepackage[matrix,frame,arrow]{xypic}
\usepackage[normalem]{ulem}
\usepackage{slashed}
\usepackage{dcolumn}
\usepackage{tabularx}
\usepackage{array}
\usepackage{adjustbox}
\usepackage{amsopn}
\usepackage{color}
\usepackage[usenames,dvipsnames,svgnames,table]{xcolor}
\usepackage[english]{babel}
\usepackage{verbatim}
\definecolor{darkblue}{rgb}{0.0,0.0,0.3}
\usepackage[colorlinks=true,
            linkcolor=red,
            urlcolor= darkblue,
            citecolor=blue]{hyperref}

\usepackage{cleveref}

\newcommand{\bea}{\begin{eqnarray}}
\newcommand{\eea}{\end{eqnarray}}

\begin{document}


\title{Dynamical one-from-many quantum metrology: Sum rule and matrix-free precision bound}

\author{Ziyu Xie}
\affiliation{Department of Physics, Institute for Quantum Science and Technology, Shanghai Key Laboratory of High Temperature Superconductors, International Center of Quantum and Molecular Structures, Shanghai University, Shanghai, 200444, China}
\author{Junjie Liu}
\email{jj\_liu@shu.edu.cn}
\affiliation{Department of Physics, Institute for Quantum Science and Technology, Shanghai Key Laboratory of High Temperature Superconductors, International Center of Quantum and Molecular Structures, Shanghai University, Shanghai, 200444, China}

\begin{abstract}
Practical quantum single-parameter estimation is rarely a pristine task; it almost invariably involves nuisance parameters, casting it as a one-from-many problem. Existing approaches to this problem rely on multi-parameter metrology, reducing the matrix quantum Cram\'er-Rao bound to obtain scalar quantum precision limits. However, these methods are often hampered by the demanding inversion of the quantum Fisher information (QFI) matrix and the requisite choice of a weight matrix, and they break down when the QFI matrix becomes singular. Here, we show that for dynamical one-from-many estimation, a previously overlooked sum rule connecting the QFI about all model parameters to the QFI about time necessitates including the latter to consider an augmented QFI matrix while simultaneously rendering it inherently singular--precisely the scenario where conventional approaches fail. To meet this challenge, we derive a tight, matrix-free quantum precision bound that involves only scalar quantities, offers broad applicability, and subsumes existing results as special cases. Validated in both unitary and noisy settings, our findings provide a refined operational framework for practical quantum single-parameter metrology.
\end{abstract}

\date{\today}
\maketitle

\begin{figure}[b!]
 \centering
\includegraphics[width=1\columnwidth]{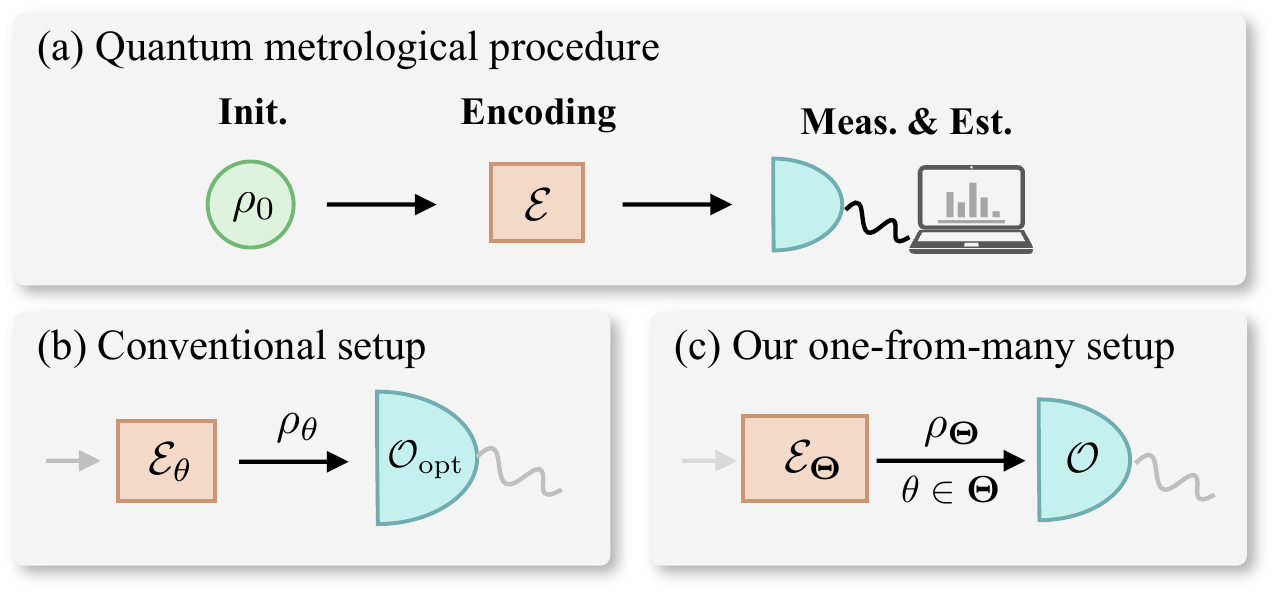} 
 \caption{(a) Standard protocol for quantum parameter estimation. (b) Conventional setup: a target-parameter parameterized encoding channel $\mathcal{E}_{\theta}$ with access to the optimal observable $\mathcal{O}_{\rm{opt}}$, for which the QCRB provides the fundamental precision limit. (c) Our one-from-many setup: a multi-parameter dynamical encoding channel $\mathcal{E}_{\boldsymbol{\Theta}}$ parameterized by a parameter set $\boldsymbol{\Theta}$, where the target parameter $\theta\in\boldsymbol{\Theta}$. We derive a tight, matrix-free precision bound [cf. Eq. (\ref{eq:lower_bound})] for estimating $\theta$, applicable to an arbitrary observable $\mathcal{O}$.}
\protect\label{fig:setup}
\end{figure}

{\it Introduction.--}Quantum metrology seeks to estimate physical parameters with a precision that surpasses classical limits~\cite{Giovannetti.04.S,Giovannetti.06.PRL,Giovannetti.11.NP,Degen.17.RMP,Pezze.18.RMP,Braun.18.RMP,Montenegro.25.PR,Sidhu.20.AVSQ,Paris.09.IJ}. A typical protocol involves initialization, parameter encoding, measurement, and estimation [Fig.~\ref{fig:setup} (a)]. In ideal scenarios of single-parameter estimation where optimal measurements are accessible, the quantum Cram\'er–Rao bound (QCRB)~\cite{Helstrom.68.IEEE,Braunstein.94.PRL,Braunstein.96.AP} dictates the ultimate precision limit, determined solely by the quantum Fisher information (QFI) $\mathcal{F}_{\theta}$ of the target parameter $\theta$~\cite{Braunstein.94.PRL,Wang.19.JPA}.
Consequently, the QFI is widely regarded as the central figure of merit~\cite{Giovannetti.04.S,Giovannetti.06.PRL,Giovannetti.11.NP,Degen.17.RMP,Pezze.18.RMP,Braun.18.RMP,Montenegro.25.PR,Sidhu.20.AVSQ,Braunstein.94.PRL,Wang.19.JPA,Helstrom.68.IEEE,Braunstein.96.AP,Scandi.25.RPP,Paris.09.IJ}, with theoretical efforts focusing heavily on its properties under saturable conditions~\cite{Boixo.07.PRL,Pang.14.PRA,Ding.23.PRL,Puig.25.PRXQ,Abiuso.25.PRL}. However, practical implementations are inevitably constrained by finite resources and restricted measurement capabilities~\cite{Kaufman.16.S,Polkovnikov.11.RMP,Trotzky.12.NP,Bernien.17.N}, which prevent saturation of the QCRB. This gap has motivated the search for tighter, more realistic bounds that generalize the QCRB~\cite{Seveso.17.PRA,Rubio.19.NJP,ZhangD.20.PRR,Len.22.NC,Gessner.23.PRL,Hervas.25.PRA,Hervas.25.PRL,Meyer.25.PRL} (see a recent review~\cite{Montenegro.25.PR} and references therein).

Standard quantum single-parameter metrology framework generally assumes that the encoding process is parameterized exclusively by the target parameter $\theta$~\cite{Boixo.07.PRL,Pang.14.PRA,ChenH.24.PRL,Das.25.PRA,Mann.25.PRXQ,Gorecki.25.PRXQ} [see Fig.~\ref{fig:setup} (b)]. However, realistic scenarios--ranging from the characterization of quantum many-body probes~\cite{Eldredge.18.PRA,Puig.25.PRXQ,Montenegro.25.PR,ShiH.24.PRL,Pintos.24.PRL} to metrology under environmental noises~\cite{Alipour.14.PRL,Sekatski.17.PRX,WanK.22.PRR,Rossi.20.PRL}--inevitably involve multiple parameters. The presence of these nuisance parameters casts practical quantum single-parameter estimation as a one-from-many task~\cite{Gross.21.JPA}. Accounting for nuisance parameters typically inflates the estimation error, thereby rendering the QCRB loose~\cite{Suzuki.20.JPA}. To faithfully capture their impact, one conventionally resorts to the framework of quantum multi-parameter metrology~\cite{Eldredge.18.PRA,Yang.19.CMP,Suzuki.20.JPA,Suzuki.20.JPAa,Gross.21.JPA,Tsang.20.PRX}. Nevertheless, obtaining a scalar precision bound for the one-from-many estimation from the matrix QCRB~\cite{Wang.19.JPA,Sidhu.20.AVSQ} is fraught with challenges~\cite{Yang.19.CMP,Suzuki.20.JPA,Suzuki.20.JPAa,Tsang.20.PRX,Gross.21.JPA}. This approach not only demands computing all QFI matrix elements for matrix inversion but also necessitates specifying a weight matrix, and the resulting bound may not be attainable. More fundamentally, this approach breaks down when the QFI matrix becomes singular~\cite{Yang.25.PRL,Mihailescu.26.QST}. These complexities necessitate an alternative framework that circumvents the matrix QCRB while addressing the one-from-many problem.

Here, we develop such a general framework for dynamical one-from-many metrology. Compared with existing approaches based on the matrix QCRB~\cite{Eldredge.18.PRA,Yang.19.CMP,Suzuki.20.JPA,Suzuki.20.JPAa,Gross.21.JPA}, our framework advances the treatment in two fundamental aspects by exploiting the multi-parameter dependence of encoding processes [Fig.~\ref{fig:setup} (c)]. First, we show that the QFI about time, which is conventionally excluded from the QFI matrix in dynamical multi-parameter metrology, should be included in the precision analysis of estimating model parameters (i.e., Hamiltonian and noise parameters), and that the resulting augmented QFI matrix is inherently non-invertible, thereby necessitating a matrix-free approach for treating dynamical one-from-many metrology. These findings follow from a fundamental sum rule [Eq.~(\ref{eq:sr})] that establishes a rigid connection between the QFI matrix elements of all model parameters and the QFI with respect to time. Second, we derive a matrix-free quantum precision bound [Eq.~(\ref{eq:lower_bound})] that relies solely on scalar quantities. Notably, our bound offers versatility across diverse metrological settings with varying numbers of model parameters, encompassing both suboptimal and optimal measurement scenarios. We demonstrate that our bound naturally subsumes existing results as special cases and validate our results through analytical and numerical examples.

{\it Dynamical setup and sum rule.--}We begin with a fully parameterized dynamical encoding channel described by the quantum Lindblad master equation~\cite{Breuer.02.NULL} (setting $k_B=1$ and $\hbar=1$ hereafter)
\begin{equation}\label{eq:lindblad}
    \partial_t \rho~=~-i[H(\{\omega_j\}), \rho]+\sum_k \gamma_k \mathcal{D}[J_k] \rho.
\end{equation}
Here, $\partial_{t}\equiv \partial/\partial {t}$ denotes time derivative, $H(\{\omega_j\})$ is the system Hamiltonian, which depends linearly on a set of Hamiltonian parameters $\{\omega_j\}$~\cite{Boixo.07.PRL,Pang.14.PRA,Puig.25.PRXQ,Montenegro.25.PR,ShiH.24.PRL,Pintos.24.PRL,Eldredge.18.PRA}, and $\mathcal{D}[J_k] \rho=J_k\rho J_k^{\dagger}-\{J_k^{\dagger}J_k,\rho\}/2$ is a dissipator with damping coefficient $\gamma_k$ and jump operator $J_k$ (where $\{A,B\}=AB+BA$). This framework encompasses a broad spectrum of quantum metrological scenarios~\cite{Degen.17.RMP,Pezze.18.RMP,Braun.18.RMP,Montenegro.25.PR,Sidhu.20.AVSQ,Paris.09.IJ,Alipour.14.PRL,Gammelmark.14.PRL,Sekatski.17.PRX,Albarelli.18.Q,WanK.22.PRR,Rossi.20.PRL,ChenH.24.PRL,Das.25.PRA,Mann.25.PRXQ,Gorecki.25.PRXQ,Mattes.25.PRL,Huelga.97.PRL}, including unitary encoding as the special case of $\{\gamma_k=0\}$. We denote $\{\theta_n\}=\{\{\omega_j\},\{\gamma_k\}\}$ as the set of model parameters, where the integer index $1\le n \le N$ with $N$ the total number of model parameters.

We observe that Eq. (\ref{eq:lindblad}) remains invariant under the scaling transformation $t\to t/\xi$ and $\{\theta_n\to\xi\theta_n\}$ for an arbitrary non-zero real factor $\xi$. This scaling invariance dictates a set of sum rules that connect the QFI about time $t$ and those of the model parameters $\{\theta_n\}$ (see Appendix A of the {\it End matter} for the full derivation)
\begin{eqnarray}
 t \mathcal{F}_{\theta_n t} &=& \sum_{m=1}^N \theta_m \mathcal{F}_{\theta_n \theta_m},\nonumber\\ 
 t \mathcal{F}_{t} &=& \sum_{n=1}^N \theta_n \mathcal{F}_{ \theta_n t}. \label{eq:sss}
\end{eqnarray}
Here, $\mathcal{F}_{xy} \equiv \mathrm{Tr}[\rho \{L_x, L_y\}]/2=\mathcal{F}_{yx}$ represents the QFI matrix elements ~\cite{Wang.19.JPA,Sidhu.20.AVSQ,Paris.09.IJ}, with $L_{x}$ ($L_y$) the symmetric-logarithmic derivative (SLD) about the parameter $x$ ($y$). We further denote $\mathcal{F}_{x}\equiv \mathcal{F}_{xx}$ as the standard QFI about parameter $x$. Combining these relations in Eq.~(\ref{eq:sss}), we obtain a remarkably compact sum rule that links all diagonal and off-diagonal elements of the QFI matrix about $\{\theta_n\}$ to the QFI with respect to time $t$, 
\begin{equation}\label{eq:sr}
    t^2 \mathcal{F}_t~=~\sum_{n=1}^N \theta_n^2 \mathcal{F}_{\theta_n} + \sum_{\substack{m, n=1  \\ m\neq n}}^N \theta_n \theta_m \mathcal{F}_{\theta_n \theta_m}.
\end{equation}
Crucially, as shown in the Supplemental Material (SM)~\cite{SM}, these sum rules extend to systems governed by generalized time-local master equations~\cite{Hall.14.PRA,Breuer.16.RMP,Vega.17.RMP}, ensuring their validity across both Markovian and non-Markovian regimes.

These sum rules, which constitute our first main result, carry profound consequences for dynamical metrology. On the one hand, they dictate that the QFI matrix elements involving time should be included in the precision analysis even when estimating model parameters $\{\theta_n\}$, since their associated QFI values are inherently coupled. This implies that time $t$ should be treated on an equal footing with all model parameters. We therefore introduce an augmented parameter set $\boldsymbol{\Theta} = \{t, \{\theta_n\}\}$, along with the associated augmented QFI matrix. On the other hand, Eq.~(\ref{eq:sss}) renders this augmented QFI matrix inherently singular (see Appendix B of the End Matter). This singularity precludes the direct application of standard matrix-based methods to the dynamical estimation of the target parameter $\theta\in \boldsymbol{\Theta}$~\footnote{We remark that conventional multi-parameter estimation excludes time from the parameter set, thereby largely avoiding the inherent singularity of the augmented QFI matrix. However, we argue that such a treatment is incomplete when the sum rule holds.}. How, then, can one determine the quantum precision limit for such one-from-many tasks? We provide a general matrix-free solution below.

{\it Matrix-free precision bound.--}We consider an arbitrary system observable $\mathcal{O}$, which may be suboptimal or optimal depending on experimental capabilities. The variance of the observable, $\mathrm{Var}(\mathcal{O})=\langle \bar{\mathcal{O}}^2\rangle$ with $\bar{\mathcal{O}}\equiv \mathcal{O}-\langle \mathcal{O}\rangle$ and $\langle \mathcal{O}\rangle=\mathrm{Tr}[\rho\mathcal{O}]$, satisfies the generalized uncertainty relation $\mathrm{Var}(\mathcal{O}) \geq |\Delta_x\langle\mathcal{O}\rangle|^2\mathcal{F}_x^{-1}$ for any parameter $x\in\boldsymbol{\Theta}$ (see Appendix C of the {\it End Matter} for the derivation). Here, $\Delta_x\langle\mathcal{O}\rangle\equiv \partial_x\langle \mathcal{O}\rangle -\langle \partial_x \mathcal{O}\rangle$. A central insight of our framework is that $x$ can be chosen as any parameter within the augmented set $\boldsymbol{\Theta}$, rather than being restricted to the target parameter $\theta$. This generalizes the standard QCRB derivation--which conventionally imposes $x=\theta$ (Appendix C of the {\it End Matter})--by exploiting the multi-parameter dependence of encoding channels. Without loss of generality, we assume $\partial_{\theta} \mathcal{O}=0$ to align with the QCRB formulation~\cite{Braunstein.94.PRL}. However, our framework explicitly accommodates cases where $\partial_x \mathcal{O}\neq 0$ for $x\neq \theta$~\footnote{For instance, this can be the case in which the system Hamiltonian constitutes the observable $\mathcal{O}$ and $x$ is a Hamiltonian parameter.}. 

By combining the uncertainty relation for $\mathrm{Var}(\mathcal{O})$ with the error propagation formula~\cite{Wineland.92.PRA,Mehboudi.19.JPA,Sidhu.20.AVSQ,Montenegro.25.PR,Mihailescu.26.PRXQ}, $\mathrm{Var}(\theta)=\mathrm{Var}(\mathcal{O})/\left|\partial_\theta \langle \mathcal{O}\rangle\right|^2$, where the variance $\mathrm{Var}(\theta)$ quantifies the estimation precision~\cite{Braunstein.94.PRL,Nichols.16.PRA,Macieszczak.16.PRA,Mihailescu.26.PRXQ},  we obtain a general matrix-free precision bound, $\mathrm{Var}(\theta)\ge \left|\frac{\Delta_x\langle \mathcal{O}\rangle}{\partial_{\theta}\langle\mathcal{O}\rangle}\right|^2\mathcal{F}_x^{-1}$, which is saturated by selecting the corresponding optimal observable $\bar{\mathcal{O}}_{\rm{opt}}^x \propto L_x$~(see Appendix C of the End Matter). This bound is expressed in terms of the response functions $\Delta_x\langle \mathcal{O}\rangle$, $\partial_{\theta}\langle\mathcal{O}\rangle$ and the QFI $\mathcal{F}_x$ about an arbitrary parameter $x$. Since $x$ can be varied over the augmented parameter set $\boldsymbol{\Theta}$, this flexibility yields a family of lower bounds $\left\{\left|\frac{\Delta_x\langle \mathcal{O}\rangle}{\partial_{\theta}\langle\mathcal{O}\rangle}\right|^2\mathcal{F}_x^{-1}\right\}$, labeled by $x$, on precision, which includes the standard QCRB $\mathcal{F}_{\theta}^{-1}$~\cite{Braunstein.94.PRL,Paris.09.IJ,Mattes.25.PRL} as the special case when $x=\theta$. By selecting the maximal bound from this family, we arrive at our second main result--a general matrix-free quantum precision bound
\begin{equation}\label{eq:lower_bound}
    \mathrm{Var}(\theta)~\ge~\max_{x\in\boldsymbol{\Theta}}\left\{\frac{1}{\mathcal{F}_x}\left|\frac{\Delta_x\langle \mathcal{O}\rangle}{\partial_{\theta}\langle\mathcal{O}\rangle}\right|^2\right\}.
\end{equation}
We emphasize that our matrix-free bound is universally applicable across diverse metrological settings: it applies to arbitrary dynamical encoding processes beyond Eq.~(\ref{eq:lindblad}) and accommodates flexibility in both the choice of the parameter set $\boldsymbol{\Theta}$ and the observable $\mathcal{O}$. Recall that $\bar{\mathcal{O}}_{\rm{opt}}^x$ saturates the bound labeled by $x$. Since this observable is generally suboptimal for any other parameter $y \neq x$, choosing $\bar{\mathcal{O}}_{\rm{opt}}^x$ maximizes the bound labeled by $x$ among all candidates, thereby saturating Eq.~(\ref{eq:lower_bound}). As $x \in \boldsymbol{\Theta}$ is arbitrary, saturating Eq.~(\ref{eq:lower_bound}) amounts to selecting an optimal observable from the set $\{\bar{\mathcal{O}}_{\rm{opt}}^x\}$, which simultaneously identifies the maximum bound.

{\it Related work.--}By construction, our matrix-free quantum bound is always tighter than the standard QCRB, namely, $\max_{x\in\boldsymbol{\Theta}}\left\{\frac{1}{\mathcal{F}_x}\left|\frac{\Delta_x\langle \mathcal{O}\rangle}{\partial_{\theta}\langle\mathcal{O}\rangle}\right|^2\right\}\ge\mathcal{F}_{\theta}^{-1}$. Our bound reduces to the QCRB in the following cases: (i) Optimal scenarios where one can select the optimal observable $\bar{\mathcal{O}}_{\rm{opt}}^{\theta} \propto L_\theta$ associated with the target parameter~\cite{Helstrom.68.IEEE}. In this case, we find 
$|\Delta_x\langle \mathcal{O}_{\rm{opt}}^{\theta}\rangle / \partial_\theta\langle \mathcal{O}_{\rm{opt}}^{\theta}\rangle|^2 = \mathcal{F}_{x\theta}^2\mathcal{F}_\theta^{-2}$ (Noting Eq. (\ref{eq:21}) in Appendix C). Using the Cauchy-Schwarz inequality $\mathcal{F}_{xy}^2 \leq \mathcal{F}_x \mathcal{F}_y$, we obtain $\mathcal{F}_x^{-1}\mathcal{F}_{x\theta}^2\mathcal{F}_\theta^{-2} \leq \mathcal{F}_\theta^{-1}$, ensuring the maximum is attained at $x = \theta$. This demonstrates the consistency of our precision bound with the QCRB in the optimal measurement scenario~\footnote{We note that the optimal observable $\mathcal{O}_{\rm{opt}}^{\theta}$ depends on other parameters in the set $\boldsymbol{\Theta}$. The ability to implement an optimal measurement implicitly assumes knowledge of other parameters, thereby reducing the one-from-many estimation problem to the conventional single-parameter estimation problem. One can thus understand why the QCRB can still serve as the tight, saturated bound in this case.}. (ii) Scenarios where the SLDs satisfy $L_{x}=k_xL_{\theta}$ with $k_x$ a coefficient. This proportionality directly implies $\mathcal{F}_x = \langle L_x^2\rangle = k_x^2 \mathcal{F}_\theta$ and $|\Delta_x\langle \mathcal{O}\rangle / \partial_\theta\langle \mathcal{O}\rangle|^2 = k_x^2$, such that each term in the set of bounds independently reduces to the QCRB for any observable. Notably, such proportionality naturally occurs in conventional dynamical single-parameter estimation settings~\cite{Boixo.07.PRL,Pang.14.PRA,ChenH.24.PRL,Das.25.PRA,Mann.25.PRXQ,Gorecki.25.PRXQ} where the augmented parameter set is assumed to be $\boldsymbol{\Theta}=\{t,\theta\}$~(see Eq.~(\ref{eq:relation_SLD})). A standard example--estimating a single multiplicative factor of a Hamiltonian under unitary encoding~\cite{Boixo.07.PRL,Pang.14.PRA}--falls into this class; another illustrative cases are provided in Appendix D and the SM~\cite{SM}.

Refs.~\cite{Eldredge.18.PRA,Gross.21.JPA} explored a variant of the one-from-many problem, where the estimation target is a linear combination of model parameters, $q=\sum_n \alpha_{\theta_n} \theta_n$, with 
$\alpha_{\theta_n}\in [-1,1]$ representing real weights. For this task, Ref.~\cite{Eldredge.18.PRA} employed the matrix QCRB about $\{\theta_n\}$ to derive a precision bound in the absence of control Hamiltonian
\begin{equation}\label{eq:q_bound}
    \mathrm{Var}(q)~\ge~\max_{x\in\{\theta_n\}}\left\{\frac{\alpha_x^2}{\mathcal{F}_x}\right\}.
\end{equation}
Notably, in the limit where only one weight  $\alpha_{\theta_i}=1$ is non-zero, Eq. (\ref{eq:q_bound}) reduces exactly to the QCRB about $\theta_i$. By extending our framework to the estimation of $q$, we analytically show that our bound Eq. (\ref{eq:lower_bound})--empowered by our sum rule--naturally recovers Eq. (\ref{eq:q_bound}) (see details in Appendix D of the {\it End Matter}). This confirms that even the estimation of arbitrary linear parameter functions can be rigorously addressed without referring to the matrix QCRB.

\begin{figure}[t!]
 \centering
 \includegraphics[width=1\columnwidth]{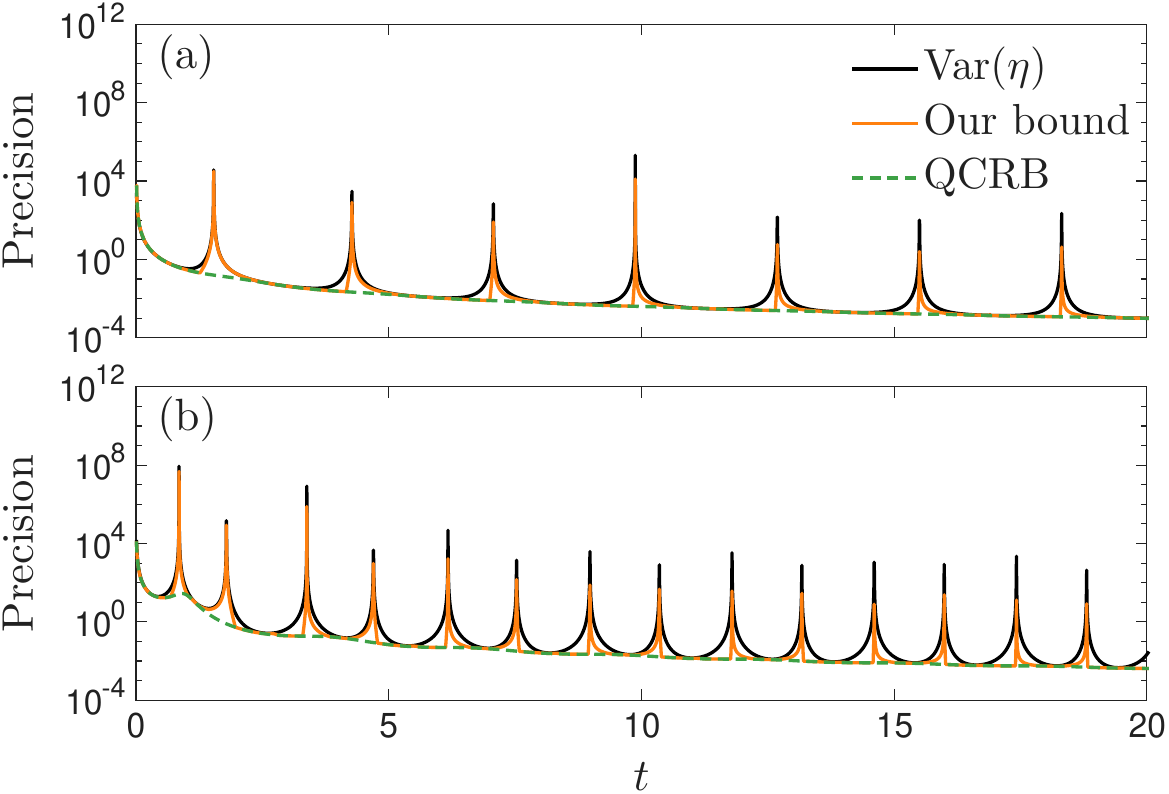} 
 \caption{Time evolution of the actual variance $\mathrm{Var}(\eta)$ (black solid line), our bound (orange solid line), and the QCRB $\mathcal{F}_{\eta}^{-1}$ (green dashed line) for the unitary model under different initial states: (a) $\rho(0) = \ket{g}\bra{g}$ and (b) $\rho(0) =0.6\ket{e}\bra{e}+ 0.4\ket{g}\bra{g}+(0.4+0.2i)\ket{e}\bra{g}+(0.4-0.2i)\ket{g}\bra{e}$ with $|g\rangle$ ($|e\rangle$) the ground (excited) state of $\sigma_z$. Parameters are $\omega = 0.5$ and $\eta = 1$.}
 \label{fig:unitary_model}
\end{figure}

{\it Example I: Unitary model.--} We consider a Hamiltonian $H(\omega, \eta)=\omega \sigma_z+\eta\sigma_x$~\cite{Abiuso.25.PRL} which generates a unitary encoding $\rho=\mathcal{U}(t)\rho_0\mathcal{U}^{\dagger}(t)$ with $\mathcal{U}(t)=\exp[-it H(\omega, \eta)]$. In this case, the parameter set $\boldsymbol{\Theta}=\{t,\omega, \eta\}$. The dynamics of this model can be solved analytically, and we show that the sum rule Eq. (\ref{eq:sr}) takes the specific form (see details in SM~\cite{SM})
\begin{equation}
   t^2\mathcal{F}_t~=~\omega^2 \mathcal{F}_{\omega} +\eta^2\mathcal{F}_\eta + \omega\eta (\mathcal{F}_{\omega \eta}+\mathcal{F}_{\eta\omega}),
\end{equation}
which holds for arbitrary initial states. 

To illustrate the utility of Eq.~(\ref{eq:lower_bound}), we consider an experimentally accessible spin observable $\sigma_z$. Figure~\ref{fig:unitary_model} presents numerical results for the variance and bounds for estimating $\theta = \eta$ for different initial states. The variance $\mathrm{Var}(\eta)$ is computed via the error propagation formula, and at each time the tight bound is identified as the maximum among the bounds in Eq.~(\ref{eq:lower_bound}). As shown in Figs.~\ref{fig:unitary_model} (a) and (b), our tight bound accurately captures the peaks of $\mathrm{Var}(\eta)$, which exhibit significant variations in magnitude, in sharp contrast to the nearly monotonically decaying QCRB $\mathcal{F}_{\eta}^{-1}$. Interestingly, we also observe that the QCRB remains tight during certain time intervals; our bound captures this behavior as well, since in such cases the QCRB simply emerges as the maximum in Eq.~(\ref{eq:lower_bound}). This example conveys a key message: for practical quantum single-parameter estimation, Eq.~(\ref{eq:lower_bound}) provides a systematic strategy to identify the tight precision limit by surveying all candidate bounds, rather than relying solely on the QCRB.

{\it Example II: Quantum thermometry.--}We then turn to noisy metrology, presenting two examples here and providing an analytically solvable spontaneous emission model in the SM~\cite{SM} for interested readers. We first focus on a qubit-based quantum thermometry that is experimentally feasible~\cite{Mehboudi.19.JPA,Kuffer.25.PRXQ}, with dynamics governed by the Lindblad master equation~\cite{Xie.26.A}
\begin{equation}\label{eq:thermemetry_lindblad}
    \partial_t \rho = - i \left[H, \rho \right] + \gamma_{+} \mathcal{D}[\sigma_{+}]\rho + \gamma_{-} \mathcal{D}[\sigma_{-}]\rho + \gamma_{0} \mathcal{D}[\sigma_{z}]\rho.
\end{equation}
Here, $H=\frac{\omega}{2}\sigma_z$ is the system Hamiltonian. The first two dissipators model the interaction with a thermal bath at temperature $T$ (the target parameter $\theta$), where the damping coefficients are $\gamma_+ = \gamma N$, $\gamma_- = \gamma (N+1)$ and $N = 1/(e^{\omega/T}-1)$. The last dissipator describes pure-dephasing with strength $\gamma_0$. The evolution is parameterized by the parameter set $\boldsymbol{\Theta}=\{t,\omega,\gamma,\gamma_0,T\}$. In the long time limit, the system approaches a thermal equilibrium state $\rho_T=e^{-H/T}/\mathrm{Tr}[e^{-H/T}]$. For this model, the sum rule Eq. (\ref{eq:sr}) simplifies to the specific form (see details in SM~\cite{SM})
\begin{equation}
    t^2 \mathcal{F}_t~=~\gamma^2 \mathcal{F}_{\gamma}+\gamma_0^2 \mathcal{F}_{\gamma_0}+ T^2 \mathcal{F}_{T}+\omega^2 \mathcal{F}_{\omega}+2\gamma \gamma_0 \mathcal{F}_{\gamma \gamma_0}+2T\omega \mathcal{F}_{T \omega},  
\end{equation}
where we have used $\mathcal{F}_{xy}=\mathcal{F}_{yx}$ to combine cross terms.

The system Hamiltonian $H$ is the optimal observable that saturates the QCRB $\mathcal{F}_T^{-1}$ at thermal equilibrium~\cite{Hovhannisyan.21.PRXQ}. Out of equilibrium, this optimality is generally expected to be lost. However, a quantitative characterization of this sub-optimality has remained absent. Eq. (\ref{eq:lower_bound}) provides the desired tool for this task: applying it to the present model while still using the system Hamiltonian as the observable, we find the tight bound can be cast into the following compact form (see details in SM~\cite{SM}; time dependence is suppressed)
\begin{equation}\label{eq:m3_bound}
    \max_{x \in \boldsymbol{\Theta}} \left
    \{\frac{1}{\mathcal{F}_x}\left| \frac{\Delta_x \langle H\rangle}{\partial_T \langle H\rangle}\right|^2\right\}~=~\frac{1}{\mathcal{F}_T + \mathcal{R}}.
\end{equation}
Here, $\mathcal{R}=\min \left\{\mathbb{E},0\right\}$ which takes the minimum among $\mathbb{E}\equiv\mathcal{F}_\gamma \left|\partial_T \langle H\rangle/\partial_\gamma \langle H\rangle\right|^2 - \mathcal{F}_T$ and $0$. It follows that the corresponding QCRB $\mathcal{F}_T^{-1}$ is no longer the tight bound whenever $\mathbb{E}<0$. 

\begin{figure}[t!]
 \centering
\includegraphics[width=1\columnwidth]{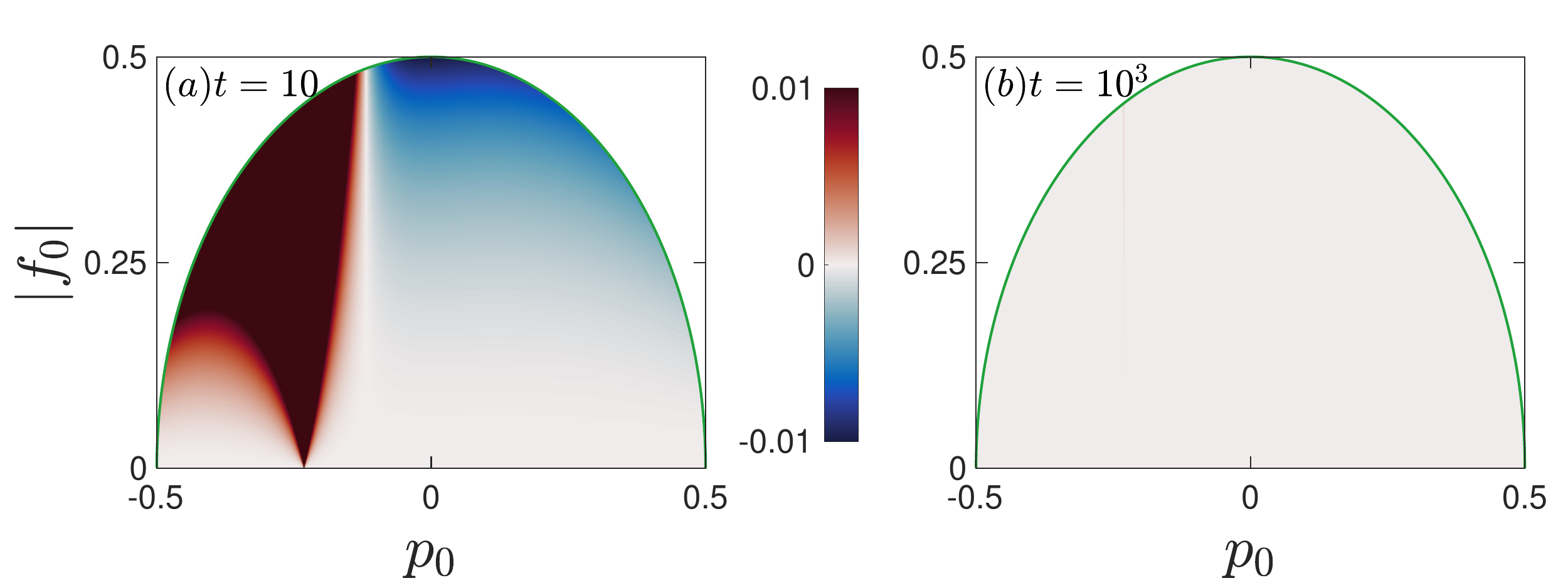} 
 \caption{The quantity $\mathbb{E}(t)$ evaluated at (a) $t=10$ and (b) $t=10^3$ for all physically allowed initial states. Initial states are parameterized by the Bloch vector $\bm{r}=(2\mathrm{Re} f_0,-2\mathrm{Im} f_0,2p_0)$ with $|p_0|\le 0.5$ and $p_0^2+|f_0|^2\le0.25$ to ensure a physical density matrix; The equality $p_0^2+|f_0|^2=0.25$ gives the green boundaries shown in the plots. Parameters are $\gamma = 0.01$, $\gamma_0 = 0.01$, $\omega =1$ and $T = 1$.}
\protect\label{fig:m3}
\end{figure}

In Fig.~\ref{fig:m3}, we plot the magnitude of $\mathbb{E}(t)$ (whose analytical expression is provided in the SM~\cite{SM}) over the full space of initial states. At short times [Fig.~\ref{fig:m3} (a)], $\mathbb{E}$ exhibits a strong dependence on initial states. In regimes where $\mathbb{E}\ge 0$, the QCRB $\mathcal{F}_T^{-1}$ remains the tight bound even out of equilibrium. However, there also exist regimes--especially for coherent initial states with large off-diagonal elements and nearly vanishing $p_0$--where $\mathbb{E}<0$. In these regions, the bound $\left| \partial_{\gamma} \langle H\rangle/\partial_T \langle H\rangle\right|^2\mathcal{F}_{\gamma}^{-1}$ becomes tighter than the QCRB. Nevertheless, at long times [Fig.~\ref{fig:m3} (b)], $\mathbb{E}$ uniformly approaches zero, and the QCRB becomes universally tight, consistent with its equilibrium optimality in the long-time limit. Our analysis therefore reveals that the precision of nonequilibrium thermometry, when assessed using the QCRB, critically depends on the initial states.

{\it Example III: Two-qubit model.--} We remark that our framework can be efficiently implemented numerically, with details provided in the SM~\cite{SM}. As a demonstration, we consider an open two-qubit model subject to local spontaneous emission, which lacks an analytical treatment. The evolution of the system state is governed by the Lindblad master equation
\begin{equation}\label{eq:master_two}
    \partial_t \rho = -i[H_2, \rho] + \gamma \left( \mathcal{D}[\sigma_-^{(1)}]\rho + \mathcal{D}[\sigma_-^{(2)}]\rho \right),
\end{equation}
with the system Hamiltonian
\[
H_2 = \omega (\sigma_z^{(1)} + \sigma_z^{(2)}) + \lambda \, \sigma_x^{(1)}\sigma_x^{(2)},
\]
where $\omega$ is the local energy gap (assumed identical for both qubits), $\lambda$ is the qubit-qubit coupling strength, $\sigma_{x,z,-}^{(j)}$ denote the usual spin operators for the $j$th qubit, and $\gamma$ is the spontaneous emission rate. For this model, the complete parameter set is $\boldsymbol{\Theta} = \{t, \omega, \lambda, \gamma\}$. We demonstrate the utility of our framework by estimating the qubit-qubit coupling strength $\theta = \lambda$.

\begin{figure}[t!]
 \centering
 \includegraphics[width=1\columnwidth]{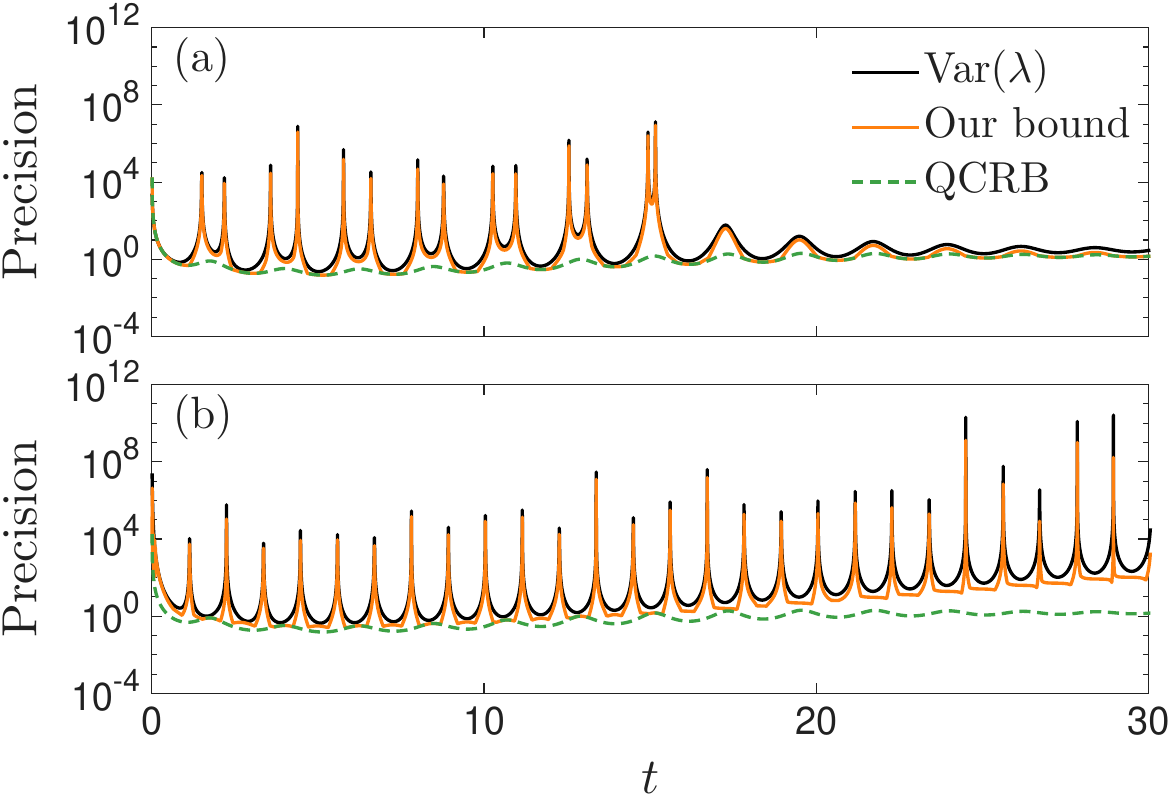} 
 \caption{Time evolution of the actual variance $\mathrm{Var}(\lambda)$ (black solid line), our bound (orange solid line), and the QCRB $\mathcal{F}_{\lambda}^{-1}$ (green dashed line) for the two-qubit model with varying observables: (a) $\mathcal{O}=\omega \sigma_z^{(1)}$ and (b) $\mathcal{O}=\sigma_x^{(1)}\sigma_x^{(2)}$. The initial state is $\rho(0) = \ket{gg}\bra{gg}$ with $|g\rangle$ the ground state of $\sigma_z^{(1),(2)}$. Parameters are $\omega = 0.5$, $\lambda = 1$, and $\gamma = 0.2$.}
 \label{fig:Numerical_model}
\end{figure}

In Fig.~\ref{fig:Numerical_model}, we compare the actual variance $\mathrm{Var}(\lambda)$, obtained via the error propagation formula, with our bound from Eq.~(\ref{eq:lower_bound}) and the QCRB $\mathcal{F}_{\lambda}^{-1}$ for varying observables. Numerical details for getting our bound are provided in the SM~\cite{SM}. As shown in Fig.~\ref{fig:Numerical_model} (a) and (b), our bound consistently provides a tighter constraint on $\mathrm{Var}(\lambda)$ than the QCRB throughout the evolution. Notably, it accurately captures the significant variations and peak structure of $\mathrm{Var}(\lambda)$ across different observables, whereas the QCRB remains small and fails to capture these features. This comparison further highlights the utility of our scalar bound in Eq.~(\ref{eq:lower_bound}) for addressing complex one-from-many estimation tasks.

{\it Conclusion.--}We have developed a general and practical  matrix-free framework for dynamical one-from-many quantum metrology. This framework is validated through a set of well-studied models, clearly demonstrating a sum rule for the QFI and showing that Eq. (\ref{eq:lower_bound}) consistently provides a tight bound across different scenarios. Our results essentially uncover previously unrecognized aspects of single-parameter precision assessment in the presence of other parameters, shifting the analysis from the matrix QCRB to a family of scalar bounds. We emphasize that the number of bounds in Eq. (\ref{eq:lower_bound}) corresponds exactly to the number of parameters in the set $\boldsymbol{\Theta}$ that fully specify the dynamical encoding channel, and does not scale exponentially with the system dimension. Consequently, applying our framework to complex systems--where analytical solutions are typically intractable--does not incur an exponential overhead. We have also shown that a numerical implementation of our framework is feasible, further facilitating the treatment of such complex systems.

{\it Acknowledgment.--}
We thank Tomi Ohtsuki for insightful discussions. We also thank anonymous referees for constructive criticism that helped to improve clarity of this manuscript. This work is supported by the National Natural Science Foundation of China (Grant No. 12205179) and the Shanghai Science and Technology Innovation Action Plan (Grant No. 24LZ1400800).


%

\onecolumngrid

\begin{center}
    {\Large \textbf{End Matter}}
\end{center}

\twocolumngrid

\setcounter{equation}{0}  
\renewcommand{\theequation}{A\arabic{equation}}

{\it Appendix A: Derivation details of sum rules.--}Applying the scaling transformation $t \to t' = t/\xi$ and $\{\theta_n \to \theta'_n = \xi \theta_n\}$, with $\xi$ an arbitrary nonzero factor, to Eq.~(\ref{eq:lindblad}) of the main text yields
\begin{equation}\label{eq:transformation_lindblad}
    \xi \frac{\partial \rho'}{\partial t} = -i \xi \sum_j \omega_j [H_j, \rho'] + \xi \sum_k \gamma_k \mathcal{D}[J_k]\rho',
\end{equation}
where $\rho' \equiv \rho(t', \{\theta_n'\})$ and $H(\{\omega_j\}) = \sum_j \omega_j H_j$. Comparing the above equation with Eq.~(\ref{eq:lindblad}) in the main text, we find that the density matrix is invariant under the scaling transformation,
\begin{equation}
    \rho(t,\{\theta_n\}) = \rho \left( \frac{t}{\xi}, \{ \xi \theta_n \} \right).
\end{equation}
Since $\xi$ is arbitrary, this invariance directly implies $\mathrm{d}\rho / \mathrm{d}\xi = 0$. Evaluating this total derivative at $\xi = 1$ gives
\begin{equation}
    \begin{aligned}
    \left. \frac{\mathrm{d}}{\mathrm{d}\xi} \rho\left(t', \{ \theta_n' \}\right) \right|_{\xi=1}
    &= \left. \left( -\frac{t}{\xi^2} \partial_{t'} \rho + \sum_{n=1}^N \theta_n \partial_{\theta_n'} \rho \right) \right|_{\xi=1} \\
    &= -t \, \partial_t \rho + \sum_{n=1}^N \theta_n \partial_{\theta_n} \rho = 0,
    \end{aligned}
\end{equation}
where $N$ is the total number of system and noise parameters. Thus, we obtain
\begin{equation}\label{eq:relation_derivative}
    t \, \partial_t \rho = \sum_{n=1}^N \theta_n \, \partial_{\theta_n} \rho,
\end{equation}

To apply Eq.~(\ref{eq:relation_derivative}) in quantum metrology, we recall the general expression for the symmetric logarithmic derivative $L_x$ with respect to an arbitrary parameter $x \in \{t, \{\theta_n\}\}$~\cite{Wang.19.JPA}:
\begin{equation}
    L_x = \sum_{k,l} \frac{2 \langle \psi_k | \partial_x \rho | \psi_l \rangle}{\mu_k + \mu_l} |\psi_k \rangle \langle \psi_l|,
\end{equation}
where $\{|\psi_k\rangle\}$ and $\{\mu_k\}$ are the eigenstates and eigenvalues of $\rho$. Substituting Eq.~(\ref{eq:relation_derivative}) into the above expression yields a linear relation among the SLD operators:
\begin{equation}\label{eq:relation_SLD}
    t L_t = \sum_{n=1}^N \theta_n L_{\theta_n}.
\end{equation}
This equation directly implies a relation among anticommutators: $\{tL_t, L_x\} = \{ \sum_{n=1}^N \theta_n L_{\theta_n}, L_x \}$. Using the QFI matrix elements $\mathcal{F}_{xy} = \mathrm{Tr}[\rho \{L_x, L_y\}]/2$ with $\mathcal{F}_{xy} = \mathcal{F}_{yx}$~\cite{Wang.19.JPA}, we then obtain
\begin{equation}\label{eq:s10}
    t \mathcal{F}_{t x} = \sum_{n=1}^N \theta_n \mathcal{F}_{\theta_n x}.
\end{equation}
Taking $x = t$ or $x = \theta_m \in \{\theta_n\}$, we recover the two sum rules in Eq.~(\ref{eq:sss}) of the main text. Equation~(\ref{eq:sr}) then follows directly by combining them.

{\it Appendix B: Singularity of the augmented QFI matrix.--}For clarity, we first illustrate the singularity of the augmented QFI matrix for a specific augmented parameter set $\boldsymbol{\Theta} = \{t, \theta_1, \theta_2\}$. A necessary and sufficient condition for a matrix to be singular is the existence of a nonzero vector that the matrix maps to zero. Using Eq.~(\ref{eq:sss}) in the main text, we identify the vector
\begin{equation}
    \mathbf{v} = [-t, \theta_1, \theta_2]^T,
\end{equation}
which satisfies
\begin{equation}
    \begin{bmatrix}
        \mathcal{F}_t & \mathcal{F}_{t\theta_1} & \mathcal{F}_{t\theta_2} \\
        \mathcal{F}_{\theta_1 t} & \mathcal{F}_{\theta_1} & \mathcal{F}_{\theta_1\theta_2} \\
        \mathcal{F}_{\theta_2 t} & \mathcal{F}_{\theta_2\theta_1} & \mathcal{F}_{\theta_2}
    \end{bmatrix}
    \begin{bmatrix}
        -t \\ \theta_1 \\ \theta_2
    \end{bmatrix}
    =
    \begin{bmatrix}
        0 \\ 0 \\ 0
    \end{bmatrix}.
\end{equation}
Here, the vanishing of each row follows directly from the sum rules for the elements of the augmented QFI matrix. Thus, the sum rules prove that the augmented QFI matrix is singular.

The generalization to an arbitrary augmented parameter set $\boldsymbol{\Theta}' = \{t, \theta_1, \dots, \theta_N\}$ is straightforward. Consider the extended vector
\begin{equation}
    \mathbf{v} = [-t, \theta_1, \dots, \theta_N]^T,
\end{equation}
which satisfies
\begin{equation}
    \mathcal{F}_Q \cdot \mathbf{v} = \mathbf{0},
\end{equation}
where $\mathcal{F}_Q$ is the augmented QFI matrix associated with $\boldsymbol{\Theta}'$ and $\mathbf{0}$ is the $(N+1)$-dimensional zero vector. This rigorously demonstrates the singularity of the augmented QFI matrix.

{\it Appendix C: Derivation of the QCRB revisited and derivation of the matrix-free precision bound.--}We begin by revisiting the standard derivation of the quantum Cram\'er-Rao bound (QCRB) for single-parameter estimation. Consider a quantum state $\rho_\theta$ parameterized solely by the target parameter $\theta$, and let $\mathcal{O}$ be an arbitrary observable. The expectation value $\langle \mathcal{O} \rangle = \mathrm{Tr}[\rho_\theta \mathcal{O}]$ then carries information about $\theta$. In the conventional single-parameter setting, the observable $\mathcal{O}$ is assumed to be independent of the parameter, so that $\partial_\theta \mathcal{O} = 0$~\cite{Braunstein.94.PRL}. We thus have
\begin{equation}\label{22}
    \partial_\theta \langle \mathcal{O} \rangle = \mathrm{Tr}[\partial_\theta \rho_\theta \, \mathcal{O}]
    = \frac{1}{2} \langle \{ L_\theta, \mathcal{O} \} \rangle,
\end{equation}
where $L_\theta$ is the symmetric logarithmic derivative (SLD) associated with $\theta$, defined by $\partial_\theta \rho = \{L_\theta, \rho_\theta\}/2$, with $\{A,B\}=AB+BA$ denoting the anticommutator. Using the property $\langle L_\theta \rangle = \mathrm{Tr}[\partial_\theta \rho_\theta] = 0$, we may shift both operators by their mean values and rewrite the above equation as
\begin{equation}\label{eq:21}
    \partial_\theta \langle \mathcal{O} \rangle = \frac{1}{2} \langle \{ \bar{L}_\theta, \bar{\mathcal{O}} \} \rangle,
\end{equation}
where $\bar{A} = A - \langle A \rangle$ denotes the centered operator for any observable $A$. Applying the Cauchy-Schwarz inequality~\cite{Landi.24.PRXQ,Vo.22.JPA} yields
\begin{equation}\label{eq:23}
    \langle \bar{L}_\theta^2 \rangle \langle \bar{\mathcal{O}}^2 \rangle
    \ge \left| \frac{1}{2} \langle \{ \bar{L}_\theta, \bar{\mathcal{O}} \} \rangle \right|^2.
\end{equation}
The equality is attained when selecting an optimal observable $\bar{\mathcal{O}}_{\rm{opt}}\propto L_\theta$. Noting that the quantum Fisher information is $\mathcal{F}_\theta = \langle L_\theta^2 \rangle = \langle \bar{L}_\theta^2 \rangle$, that $\mathrm{Var}(\mathcal{O}) = \langle \bar{\mathcal{O}}^2 \rangle$, and using the error-propagation expression $\mathrm{Var}(\theta) = \mathrm{Var}(\mathcal{O}) / |\partial_\theta \langle \mathcal{O} \rangle|^2$, we immediately obtain the QCRB,
\begin{equation}\label{eq:24}
    \mathrm{Var}(\theta) \ge \frac{1}{\mathcal{F}_\theta}.
\end{equation}

We now generalize the above derivation to obtain a matrix-free precision bound applicable to the one-from-many estimation scenario while avoiding the QFI matrix--a main result of our study. In many practical quantum metrological tasks, one encounters a multi-parameter quantum state $\rho_{\boldsymbol{\Theta}}$ with parameter set $\boldsymbol{\Theta}$ and target parameter $\theta \in \boldsymbol{\Theta}$. This multi-parameter dependence leads to a key observation: we may consider the partial derivative $\partial_x \langle \mathcal{O} \rangle$ for any parameter $x \in \boldsymbol{\Theta}$, which can differ from $\theta$. This extension naturally generalizes the QCRB to account for nuisance parameters, recovering the standard QCRB when $x = \theta$. Noting that $\mathcal{O}$ may depend on $x$ when $x \neq \theta$, we have
\begin{equation}
    \partial_x \langle \mathcal{O} \rangle = \mathrm{Tr}[(\partial_x \rho_{\boldsymbol{\Theta}}) \, \mathcal{O}] + \langle \partial_x \mathcal{O} \rangle.
\end{equation}
For simplicity, we introduce the notation
\begin{equation}
    \Delta_x \langle \mathcal{O} \rangle \equiv \partial_x \langle \mathcal{O} \rangle - \langle \partial_x \mathcal{O} \rangle
    = \mathrm{Tr}[(\partial_x \rho_{\boldsymbol{\Theta}}) \, \mathcal{O}].
\end{equation}
Analogously to Eq.~(\ref{eq:21}), we obtain
\begin{equation}
    \Delta_x \langle \mathcal{O} \rangle = \frac{1}{2} \langle \{ \bar{L}_x, \bar{\mathcal{O}} \} \rangle,
\end{equation}
where $L_x$ is the SLD operator associated with the parameter $x$. Applying the Cauchy-Schwarz inequality to the above equation gives
\begin{equation}
    \langle \bar{L}_x^2 \rangle \langle \bar{\mathcal{O}}^2 \rangle
    \ge \left| \frac{1}{2} \langle \{ \bar{L}_x, \bar{\mathcal{O}} \} \rangle \right|^2,
\end{equation}
which directly generalizes Eq.~(\ref{eq:23}). This inequality is saturated when selecting an optimal observable $\bar{\mathcal{O}}_{\rm{opt}}^x\propto L_x$. Combining this inequality with the error-propagation expression for $\theta$, and noting that $\mathcal{F}_x = \langle L_x^2 \rangle$, we obtain a matrix-free precision bound for any parameter $x \in \boldsymbol{\Theta}$:
\begin{equation}\label{eq:29}
    \mathrm{Var}(\theta) \ge \frac{1}{\mathcal{F}_x} \left| \frac{\Delta_x \langle \mathcal{O} \rangle}{\partial_\theta \langle \mathcal{O} \rangle} \right|^2.
\end{equation}
It is evident that Eq.~(\ref{eq:29}) subsumes Eq.~(\ref{eq:24}) as the special case $x = \theta$. Selecting the maximum bound by varying $x$ over the parameter set $\boldsymbol{\Theta}$, we arrive at the tight matrix-free bound shown in Eq.~(\ref{eq:lower_bound}) of the main text.

{\it Appendix D: Recovering an existing bound with nuisance parameters.--}In this appendix, we show that our matrix-free bound in Eq.~(\ref{eq:lower_bound}) of the main text recovers an existing result based on the matrix QCRB~\cite{Eldredge.18.PRA}. Ref.~\cite{Eldredge.18.PRA} considered an invariant of one-from-many metrology: estimating a function $q$ that is a linear combination of parameters
\begin{equation}
    q = \sum_{n=1}^N \alpha_{\theta_n} \theta_n,
\end{equation}
where the coefficients $\alpha_{\theta_n} \in [-1,1]$ are real numbers and $\{\theta_n\}$ are model parameters. Using the matrix QCRB that involves the QFI matrix about $\{\theta_n\}$, the authors of Ref.~\cite{Eldredge.18.PRA} derived the following scalar lower bound on the estimation precision of $q$ in the absence of a control Hamiltonian:
\begin{equation}
    \mathrm{Var}(q) \ge \max_{x \in \{\theta_n\}} \left\{ \frac{\alpha_x^2}{\mathcal{F}_x} \right\},
\end{equation}
where $\mathcal{F}_x$ denotes the QFI with respect to the parameter $x$.

To facilitate comparison, we note that our matrix-free precision bound in Eq.~(\ref{eq:lower_bound}) can be directly generalized to the task of estimating $q$ by regarding $q$ as a single effective parameter:
\begin{equation}
    \mathrm{Var}(q) \ge \max_{x \in \boldsymbol{\Theta}} \left\{ \frac{1}{\mathcal{F}_x} \left| \frac{\Delta_x \langle \mathcal{O} \rangle}{\partial_q \langle \mathcal{O} \rangle} \right|^2 \right\},
\end{equation}
where $\boldsymbol{\Theta} = \{t, \{\theta_n\}\}$. We first observe that
\begin{equation}
    \partial_{\theta_n} \rho = \frac{\partial q}{\partial \theta_n} \partial_q \rho = \alpha_{\theta_n} \partial_q \rho.
\end{equation}
Using the relation $\partial_a \rho = \{L_a, \rho\}/2$ for an arbitrary parameter $a$, the above equation directly implies a connection among the SLDs:
\begin{equation}\label{eq:35}
    L_{\theta_n} = \alpha_{\theta_n} L_q,
\end{equation}
where $L_q$ is the SLD associated with $q$ as a single parameter.

Since Ref.~\cite{Eldredge.18.PRA} concerns dynamical metrology, combining the sum rule for the SLDs in Eq.~(\ref{eq:relation_SLD}) yields
\begin{equation}\label{eq:36}
    t L_t = \left( \sum_n \alpha_{\theta_n} \theta_n \right) L_q = q L_q.
\end{equation}
Consequently, Eqs.~(\ref{eq:35}) and~(\ref{eq:36}) imply that for every parameter $x \in \boldsymbol{\Theta}$ there exists a coefficient $k_x$ such that
\begin{equation}
    L_x = k_x L_q,
\end{equation}
with $k_t = q/t$ and $k_{\theta_n} = \alpha_{\theta_n}$. Thus, all SLDs with respect to time and model parameters are proportional to $L_q$.

We now exploit this proportionality to simplify the terms appearing in our matrix-free bound. First, the QFI with respect to any parameter $x$ satisfies
\begin{equation}
    \mathcal{F}_x = \langle L_x^2 \rangle = k_x^2 \langle L_q^2 \rangle = k_x^2 \mathcal{F}_q.
\end{equation}
Second, for an arbitrary observable $\mathcal{O}$, we have
\begin{equation}
    \Delta_x \langle \mathcal{O} \rangle = \frac{1}{2} \langle \{ \bar{L}_x, \bar{\mathcal{O}} \} \rangle
    = \frac{k_x}{2} \langle \{ \bar{L}_q, \bar{\mathcal{O}} \} \rangle
    = k_x \Delta_q \langle \mathcal{O} \rangle.
\end{equation}
Moreover, $\Delta_q \langle \mathcal{O} \rangle = \partial_q \langle \mathcal{O} \rangle$ when $\mathcal{O}$ is independent of $q$, so that
\begin{equation}
    \frac{\Delta_x \langle \mathcal{O} \rangle}{\partial_q \langle \mathcal{O} \rangle} = k_x.
\end{equation}
Combining the above results, we find
\begin{equation}
    \frac{1}{\mathcal{F}_x} \left| \frac{\Delta_x \langle \mathcal{O} \rangle}{\partial_q \langle \mathcal{O} \rangle} \right|^2
    = \frac{k_x^2}{\mathcal{F}_x}
    = \frac{1}{\mathcal{F}_q}.
\end{equation}
Noting that $k_x^2 / \mathcal{F}_x = \alpha_x^2 / \mathcal{F}_x$ for $x \in \{\theta_n\}$ and that every bound reduces to $\mathcal{F}_q^{-1}$ in this special case, we finally arrive at
\begin{equation}
    \max_{x \in \boldsymbol{\Theta}} \left\{ \frac{1}{\mathcal{F}_x} \left| \frac{\Delta_x \langle \mathcal{O} \rangle}{\partial_q \langle \mathcal{O} \rangle} \right|^2 \right\}
    = \max_{x \in \{\theta_n\}} \left\{ \frac{\alpha_x^2}{\mathcal{F}_x} \right\}
    = \frac{1}{\mathcal{F}_q}.
\end{equation}
Thus, our bound recovers the result of Ref.~\cite{Eldredge.18.PRA} without invoking the matrix QCRB.

This consistency can be illustrated with a concrete example involving an optimal setup analyzed in Ref.~\cite{Eldredge.18.PRA}. Consider $N$ qubits prepared in the Greenberger--Horne--Zeilinger state $\ket{\psi_0} = \frac{1}{\sqrt{2}} (\ket{0}^{\otimes N} + \ket{1}^{\otimes N})$. The unitary evolution is generated by
\begin{equation}
    U = \exp\left( -\frac{it}{2} \sum_{n=1}^N \alpha_{\theta_n} \theta_n \sigma_z^{(n)} \right),
\end{equation}
so that the state at time $t$ becomes
\[
\ket{\psi_t} = \frac{1}{\sqrt{2}} \left( e^{-iqt/2} \ket{0}^{\otimes N} + e^{iqt/2} \ket{1}^{\otimes N} \right).
\]
We take the parity operator $P = \otimes_{n=1}^N \sigma_x^{(n)}$~\cite{Eldredge.18.PRA} as the observable $\mathcal{O}$, whose expectation value is
\begin{equation}
    \langle P \rangle = \langle \psi_t | P | \psi_t \rangle = \cos(qt).
\end{equation}
From this, we directly obtain $\partial_q \langle P \rangle = -t \sin(qt)$, $\Delta_t \langle P \rangle = -q \sin(qt)$, and $\Delta_{\theta_n} \langle P \rangle = -\alpha_{\theta_n} t \sin(qt)$. Hence,
\begin{equation}
    \frac{\Delta_t \langle P \rangle}{\partial_q \langle P \rangle} = \frac{q}{t}, \qquad
    \frac{\Delta_{\theta_n} \langle P \rangle}{\partial_q \langle P \rangle} = \alpha_{\theta_n}.
\end{equation}
For this model, the QFI can be computed directly: $\mathcal{F}_{\theta_m \theta_n} = \alpha_{\theta_m} \alpha_{\theta_n} t^2$. Substituting this expression into the sum rule yields
\begin{equation}
    \begin{aligned}
        t^2 \mathcal{F}_t
        &= \sum_{m,n=1}^N \theta_m \theta_n \mathcal{F}_{\theta_m \theta_n} \\
        &= \left( \sum_{m=1}^N \alpha_{\theta_m} \theta_m \right)
           \left( \sum_{n=1}^N \alpha_{\theta_n} \theta_n \right) t^2 \\
        &= q^2 t^2,
    \end{aligned}
\end{equation}
which gives $\mathcal{F}_t = q^2$ and $\mathcal{F}_q = t^2$. Therefore,
\begin{equation}
    \max_{x \in \boldsymbol{\Theta}} \left\{ \frac{1}{\mathcal{F}_x} \left| \frac{\Delta_x \langle P \rangle}{\partial_q \langle P \rangle} \right|^2 \right\}
    = \max_{x \in \{\theta_n\}} \left\{ \frac{\alpha_x^2}{\mathcal{F}_x} \right\}
    = \frac{1}{\mathcal{F}_q}
    = \frac{1}{t^2}.
\end{equation}
The above Heisenberg scaling in time is a direct consequence of the optimal entangled probe state used.

\onecolumngrid

\begin{center}
    {\Large \textbf{Supplemental Material: Dynamical one-from-many quantum metrology: Sum rule and matrix-free precision bound}}
\end{center}

\renewcommand{\theequation}{S\arabic{equation}}
\renewcommand{\thefigure}{S\arabic{figure}}
\setcounter{equation}{0}  
\setcounter{figure}{0} 

This Supplemental Material contains complementary derivations and results that support the analysis in the main text. Section~I details the derivation of the sum rules for the quantum Fisher information (QFI) matrix elements extended to non-Markovian encoding processes described by a generalized time-local master equation~\cite{Hall.14.PRA,Breuer.16.RMP,Vega.17.RMP}. Section~II presents the full complementary details for the model systems studied in the main text. In Section~III, we provide an additional analytically solvable model.

\section{I.~Sum rule in the non-Markovian regime}
In this section, we show that sum rules between the elements of the quantum Fisher information (QFI) matrix hold for non-Markovian encoding dynamics described by a generalized time-local master equation~\cite{Hall.14.PRA,Breuer.16.RMP,Vega.17.RMP}
\begin{equation}\label{eq:TCL}
    \frac{\partial}{\partial t} \rho~=~\mathcal{L} (t) [\rho]~=~-i[\mathcal{K}(t),\rho] + \widetilde{\mathcal{D}}(t)[ \rho],
\end{equation}
where $\mathcal{K} (t) = \sum_j \omega_j(t) \mathcal{K}_j$ is an effective Hamiltonian with $\{\omega_j(t)\}$ a set of time-dependent parameters and $\mathcal{K}_j$ the associated Hermitian operators whose detailed forms depend on the given model, and $ \widetilde{\mathcal{D}}(t)[ \rho] = \sum_k\gamma_k(t) \mathcal{D}[J_k] \rho$ is the Lindblad dissipator with time-dependent damping coefficients $\gamma_k(t)$ and jump operators $J_k$. This time-local master equation can describe non-Markovian dynamics~\cite{Hall.14.PRA,Breuer.16.RMP,Vega.17.RMP}.

We consider the general case where the parameters $\omega_j(t)$ and $\gamma_k(t)$ exhibit power-law time dependencies, expressed as Laurent series expansions
\begin{equation}\label{eq:series_expansion}
     \omega_j(t) = \sum_{p=-\infty}^\infty \omega_{jp} t^p,~~\gamma_k(t) = \sum_{q=-\infty}^\infty \gamma_{kq} t^q.
\end{equation}
Substituting Eq. (\ref{eq:series_expansion}) into Eq. (\ref{eq:TCL}) gives the expanded form 
\begin{equation}\label{eq:TCL_series}
    \frac{\partial}{\partial t} \rho~=~-i \sum_j  \sum_{p=-\infty}^\infty \omega_{jp} t^p [ \mathcal{K}_j ,\rho] + \sum_k \sum_{q=-\infty}^\infty \gamma_{kq} t^q  \mathcal{D}[J_k] \rho
\end{equation}
We introduce a scaling transformation analogous to the Markovian case,
\begin{equation}
    t \rightarrow t' = \frac{t}{\xi},~~\{\omega_{jp}\} \rightarrow \{\omega_{jp}'\} = \{\xi^{p+1} \omega_{jp}\},~~\{\gamma_{kq}\} \rightarrow \{\gamma_{kq}'\} = \{\xi^{q+1} \gamma_{kq}\},
\end{equation}
where $\xi \neq 0$. Comparing the transformed equation with the original Eq. (\ref{eq:TCL_series}) multiplied by the same factor $\xi$ reveals an identical structure, implying the invariance
\begin{equation}
    \rho \left(t,\{\omega_{jp}\},\{\gamma_{kq}\} \right)~=~\rho \left(\frac{t}{\xi},\{ \xi^{p+1}\omega_{jp}\}, \{\xi^{q+1}\gamma_{kq}\} \right).
\end{equation}
Since $\rho$ does not depend on $\xi$, we have $\mathrm{d}\rho/\mathrm{d}\xi = 0$. Evaluating this derivative at $\xi = 1$ gives
\begin{equation}
    t \partial_t \rho~=~\sum_j \sum_{p=-\infty}^{\infty}(p+1)\omega_{jp}\partial_{\omega_{jp}}\rho+\sum_k \sum_{q=-\infty}^{\infty}(q+1)\gamma_{kq}\partial_{\gamma_{kq}}\rho.
\end{equation}

To simplify the notation, we introduce a complete set of independent parameters $\boldsymbol{\Theta}=\{t,\{\theta_{nr}\}\}$ with $\{\theta_{nr}\}$ containing all expansion coefficients $\{\omega_{jp}\}$ and $\{\gamma_{kq}\}$, where the index $n\ge 0$ labels distinct physical operators, and $r \in \mathbb{Z}$ indexes the power of time in the Laurent-series expansion. The above derivative relation then reads
\begin{equation}
    t \partial_t \rho~=~\sum_{n\ge0} \sum_{r=-\infty}^{\infty}(r+1)\theta_{nr}\partial_{\theta_{nr}}\rho.
\end{equation}
Using the definition of the symmetric logarithmic derivative (SLD), this leads to the corresponding SLD relation
\begin{equation}
    t L_t~=~\sum_{n\ge0} \sum_{r=-\infty}^{\infty}(r+1)\theta_{nr}L_{\theta_{nr}}.
\end{equation}
Taking the anti-commutator with an arbitrary $L_x$ and tracing with $\rho$ yields
\begin{equation}\label{eq:s9}
    t \mathcal{F}_{tx}~=~\sum_{n\ge0} \sum_{r=-\infty}^{\infty}(r+1)\theta_{nr}\mathcal{F}_{\theta_{nr}x}.
\end{equation}
Setting $x=t$ gives a relation involving the QFI for time, $t \mathcal{F}_{t}=\sum_{n\ge0} \sum_{r=-\infty}^{\infty}(r+1)\theta_{nr}\mathcal{F}_{\theta_{nr}t}$. Multiplying this equation by $t$ and using Eq. (\ref{eq:s9}) with $x = \theta_{nr}$ eventually produces the generalized sum rule
\begin{equation}\label{eq:TCL_sum_rule}
    t^2 \mathcal{F}_{t}~=~\sum_{m\ge0,n\ge0} \sum_{r=-\infty}^{\infty}\sum_{s=-\infty}^{\infty}(s+1)(r+1)\theta_{ms}\theta_{nr}\mathcal{F}_{\theta_{ms}\theta_{nr}}.
\end{equation}
When only the zeroth-order terms in the Laurent series are nonzero ($\theta_{nr}=0$ for all $r \neq 0$), the generalized time-local master equation reduces to the standard time-independent Lindblad master equation studied previously. In this limit, Eq.~\eqref{eq:TCL_sum_rule} collapses to the simpler Markovian sum rule derived earlier,
\begin{equation}
    t^2 \mathcal{F}_{t}~=~\sum_{m\ge0,n\ge0}\theta_{m0}\theta_{n0}\mathcal{F}_{\theta_{m0}\theta_{n0}},
\end{equation}
which coincides with Eq.~(3) in the main text. The derivation above demonstrates that the sum rule governing the QFI matrix is not limited to Markovian dynamics but holds also for non-Markovian dynamics described by a generalized time-local master equation. This broad generality stems from the scale invariance of the dynamical equation under a suitably generalized parameter transformation. The result underscores that the interdependence of QFI matrix elements--and the consequent impossibility of attaining a nonsingular QFI matrix with respect to model parameters--is a fundamental feature of quantum dynamics, independent of memory effects. 

\section{II.~Details of model systems in the main text}
This section contains detailed derivations and numerical details for the model systems discussed in the main text.

\subsection{A.~Preliminaries for analytically solvable models}\label{subsec:II.A}
To establish a foundation for subsequent derivations and to fix notations, we first derive explicit working expressions for the QFI matrix elements $\mathcal{F}_x$ and $\mathcal{F}_{xy}$ tailored to two-level systems. For such a system, the Hilbert space can be spanned by the ground state $\ket{g}$ and the excited state $\ket{e}$ of $\sigma_z$, 
\begin{equation}
    \ket{g}~=\begin{bmatrix} 0 \\ 1 \end{bmatrix},~~\ket{e}~=~\begin{bmatrix} 1 \\ 0 \end{bmatrix}.
\end{equation}
The density matrix can be conveniently parameterized as (time-dependence is suppressed for simplicity)
\begin{equation}\label{eq:rho_expression}
    \rho~=~\begin{bmatrix}
        \frac{1}{2}+p &f \\f^* &\frac{1}{2}-p
    \end{bmatrix}~=~\left(\frac{1}{2}+p\right)\ket{e}\bra{e}+\left(\frac{1}{2}-p\right)\ket{g}\bra{g}+(f)\ket{e}\bra{g}+(f^*)\ket{g}\bra{e}.
\end{equation}
where $p \in \mathbb{R}$ and $f \in \mathbb{C}$ are real and complex functions, respectively, and $f^*$ denotes the complex conjugate of $f$. This form automatically ensures Hermiticity $\rho^\dagger = \rho$ and normalization condition $\mathrm{Tr}[\rho] = 1$. For $\rho$ to represent a valid density matrix, i.e., positive semi‑definite matrix, its determinant must be non‑negative, $\det [\rho] \geq 0$, which gives $p^2+|f|^2\leq 1/4$ and $|p|\leq 1/2$. Eq. (\ref{eq:rho_expression}) is essentially equivalent to the standard Bloch‑sphere representation $\rho = (\mathbb{I}+\vec{r}\cdot\vec{\sigma})/2$, where $\vec{r}=(r_x,r_y,r_z)^T$ is the Bloch vector with $\mathbb{I}$ the $2\times 2$ identity matrix and $\vec{\sigma}=(\sigma_x,\sigma_y,\sigma_z)^T$ the vector of Pauli matrices. 

The spectral decomposition of $\rho$ reads
\begin{equation}
    \rho~=~U\Lambda U^\dagger~=~\mu_+ \ket{\psi_+}\bra{\psi_+} + \mu_- \ket{\psi_-}\bra{\psi_-}.
\end{equation}
Here, $\Lambda = \operatorname{diag}(\mu_+, \mu_-)$ contains the eigenvalues 
\begin{equation}\label{eq:eigenvalue_rho}
    \mu_\pm~=~1/2 \pm \sqrt{p^2 + |f|^2} \geq 0,
\end{equation}
with $\mu_++\mu_-=1$, and 
\begin{equation}\label{eq:UUU}
    U~=~\left[\ket{\psi_+}, \ket{\psi_-}\right]
\end{equation}
is the unitary matrix whose columns are the corresponding eigenstates
\begin{equation}\label{eq:eigenstate_rho}
    \ket{\psi_\pm}~=~\frac{1}{\sqrt{|f|^2+A_\mp^2}} \begin{bmatrix}
        -f \\ A_\mp
    \end{bmatrix},
\end{equation}
where $A_\pm=p \pm \sqrt{p^2 + |f|^2}$. In this eigenbasis, the SLD operator for a parameter $x$ takes the form
\begin{equation}\label{eq:ULx}
    L_x~=~U(U^\dagger L_x U)U^\dagger~=~ U\begin{bmatrix}
        \bra{\psi_+}\partial_x \rho\ket{\psi_+}/\mu_+ & 2\bra{\psi_+}\partial_x \rho\ket{\psi_-}\\
        2\bra{\psi_-}\partial_x \rho\ket{\psi_+} &\bra{\psi_-}\partial_x \rho\ket{\psi_-}/\mu_-
    \end{bmatrix}U^\dagger.
\end{equation}
With the above form, one can readily check that the QFI $\mathcal{F}_x = \mathrm{Tr}[\rho L_x^2]$ reduces to the well-known sum formula~\cite{Wang.19.JPA}
\begin{equation}
    \mathcal{F}_{x}~=~2\sum_{m,n} \frac{|\langle \psi_m|\partial_x \rho|\psi_n\rangle|^2}{\mu_m+\mu_n},
\end{equation}
where $\mu_m+\mu_n \neq 0$. For a two‑level system, this expression expands to
\begin{equation}
     \mathcal{F}_{x}~=~\frac{|\bra{\psi_+}\partial_x \rho\ket{\psi_+}|^2}{\mu_+} + \frac{|\bra{\psi_-}\partial_x \rho\ket{\psi_-}|^2}{\mu_-} + 2|\bra{\psi_+}\partial_x \rho\ket{\psi_-}|^2+2|\bra{\psi_-}\partial_x \rho\ket{\psi_+}|^2.
\end{equation}
Using the facts that $\mu_++\mu_- = 1$, $\bra{\psi_+}\partial_x \rho\ket{\psi_+}+\bra{\psi_-}\partial_x \rho\ket{\psi_-}=0$ and $\bra{\psi_+}\partial_x \rho\ket{\psi_-} = (\bra{\psi_-}\partial_x \rho\ket{\psi_+}) ^*$, the above expression condenses into the compact two‑term form
\begin{equation}\label{eq:QFI_compact}
     \mathcal{F}_{x}~=~\frac{|\bra{\psi_+}\partial_x \rho\ket{\psi_+}|^2}{\mu_+\mu_-} + 4|\bra{\psi_+}\partial_x \rho\ket{\psi_-}|^2.
\end{equation}
Comparing Eqs. (\ref{eq:ULx}) and (\ref{eq:QFI_compact}), we observe that the QFI $\mathcal{F}_x$ for a two-level system can also be determined as 
\begin{equation}\label{eq:Fx_simple}
   \mathcal{F}_x~=~-\det(U^\dagger L_x U)~=~-\det(L_x). 
\end{equation}
The off-diagonal QFI matrix elements are defined as $\mathcal{F}_{xy} = \mathrm{Tr}[\rho \{L_x,L_y\}]/2$~\cite{Wang.19.JPA}. In the eigenbasis of $\rho$, this becomes
\begin{equation}\label{eq:QFIM_element}
    \mathcal{F}_{xy} = \frac{\bra{\psi_+}\partial_x \rho\ket{\psi_+}\bra{\psi_+}\partial_y \rho\ket{\psi_+}}{\mu_+\mu_-} +4\mathrm{Re}[\bra{\psi_+}\partial_x \rho\ket{\psi_-} \bra{\psi_-}\partial_y \rho\ket{\psi_+}].
\end{equation}
Equations~(\ref{eq:QFI_compact}), (\ref{eq:Fx_simple}), and~(\ref{eq:QFIM_element}) provide explicit, closed-form expressions for the QFI matrix elements of a generic two-level system. These expressions serve as the starting point for deriving the specific sum rules and precision bounds for the analytically solvable Examples~I and II in the main text.


\subsection{B.~An example showing $L_{x\neq\theta}\propto L_{\theta}$ and rendering the QCRB universally tight}
Before presenting the examples discussed in the main text, we first demonstrate a limiting case in which the QCRB emerges as the tight bound among all members of Eq.~(4) of the main text, for any observable and any number of parameters in the augmented parameter set. Consider a two-level system whose state remains incoherent throughout its evolution. In this case, the derivative of the density matrix with respect to a parameter $x$ simplifies to
\begin{equation}
    \partial_x \rho~=~\begin{bmatrix}
        \partial_x p &0 \\0 &- \partial_x p
    \end{bmatrix}~=~(\partial_x p) \sigma_z.
\end{equation}  
Consequently, derivatives with respect to two arbitrary parameters $x$ and $y$ satisfy $(\partial_y p ) \partial_x \rho=(\partial_x p)\partial_y \rho$. This leads to a proportionality relation between the corresponding SLD operators
\begin{equation}
    L_x~=~\frac{\partial_x p}{\partial_y  p} L_{y}.
\end{equation}
Thus, all SLD operators are mutually proportional. In particular, $L_x = k_x L_\theta$ for the target parameter $\theta$, which implies $\mathcal{F}_{x}=k_x^2\mathcal{F}_{\theta}$. To connect bounds in Eq. (4) of the main text with the QCRB, we restrict ourselves to observables that are independent of the target parameter, $\partial_{\theta}\mathcal{O}=0$, in accordance with the conventional assumption of parameter-independent measurements~\cite{Braunstein.94.PRL}. Under the proportionality condition $L_x = k_x L_\theta$ for every parameter $x$, we obtain
\begin{equation}
     \frac{1}{\mathcal{F}_x}\left|\frac{\Delta_x\langle \mathcal{O}\rangle}{\partial_{\theta}\langle\mathcal{O}\rangle}\right|^2~=~\frac{1}{\mathcal{F}_x}\left|\frac{\partial_x\langle \mathcal{O}\rangle}{\partial_{\theta}\langle\mathcal{O}\rangle}\right|^2~=~\frac{1}{\mathcal{F}_\theta}.
\end{equation}
Eq. (4) of the main text therefore yields the same bound for every $x$, and the maximum coincides with the QCRB $\mathcal{F}_{\theta}^{-1}$. This shows that, even when suboptimal observables are employed, the QCRB remains the tight achievable precision limit in this incoherent scenario. In contrast, all examples analyzed in the main text exhibit nonzero coherence as we show below, which generically prevents the QCRB from being universally tight.

\subsection{C.~Example I:~Unitary Model}
As our first example in the main text, we consider quantum unitary metrology where the parameter encoding channel is unitary. In this case, the system density matrix $\rho$ evolves according to the von Neumann equation,
\begin{equation}
    \partial_t \rho~=~-i[H, \rho].
\end{equation}
The unitary model we consider consists of a single qubit described by a two‑parameter Hamiltonian $H(\omega, \eta) = \omega \sigma_z+\eta \sigma_x$ which has been studied in recent metrological literature~\cite{Abiuso.25.PRL}.

\subsubsection{1.~Sum rule}
To verify the sum rule for the QFI, let us check whether Eq. (A4) of the end matter is satisfied. The terms appearing in this equation are
\begin{equation}\label{eq:rho_deri}
    \partial_t\rho~=~\begin{bmatrix}
        \partial_t p_t &\partial_t f_t \\
        \partial_t f_t^* &-\partial_t p_t
    \end{bmatrix},~~
    \partial_\omega\rho~=~\begin{bmatrix}
        \partial_\omega p_t &\partial_\omega f_t \\
        \partial_\omega f_t^* &-\partial_\omega p_t
    \end{bmatrix},~~
    \partial_\eta\rho~=~\begin{bmatrix}
        \partial_\eta p_t &\partial_\eta f_t \\
        \partial_\eta f_t^* &-\partial_\eta p_t
    \end{bmatrix}.
\end{equation}
To evaluate this condition, we need the explicit analytical expressions for $p_t$ and $f_t$. For the unitary model under consideration, the time-dependent density-matrix elements $p_t$ and $f_t = \mathrm{Re}[f_t]+i\cdot\mathrm{Im}[f_t]$ can be expressed analytically as
\begin{eqnarray}\label{eq:solution_unitary}
 p_t &=& p_0-2\frac{\partial \Omega}{\partial \eta}\left[\frac{\partial \Omega}{\partial \eta} p_0-\frac{\partial \Omega}{\partial \omega}\mathrm{Re}[f_0]\right]\sin^2(\Omega t)-\frac{\partial\Omega}{\partial \eta}\mathrm{Im}[f_0]\sin(2\Omega t), \nonumber\\ 
 \mathrm{Re}[f_t] &=& \mathrm{Re}[f_0]+2\frac{\partial \Omega}{\partial \omega}\left[\frac{\partial \Omega}{\partial \eta} p_0-\frac{\partial \Omega}{\partial \omega}\mathrm{Re}[f_0]\right]\sin^2(\Omega t)+\frac{\partial\Omega}{\partial \omega}\mathrm{Im}[f_0]\sin(2\Omega t), \nonumber\\
 \mathrm{Im}[f_t] &=& \mathrm{Im}[f_0]-2\,\mathrm{Im}[f_0]\sin^2(\Omega t)+\left[\frac{\partial \Omega}{\partial \eta} p_0-\frac{\partial \Omega}{\partial \omega}\mathrm{Re}[f_0]\right]\sin(2\Omega t),
\end{eqnarray}
where $\Omega = \sqrt{\omega^2 + \eta^2}$ is the frequency factor, with $\partial \Omega/\partial \omega = \omega/\Omega$ and $\partial \Omega/\partial \eta = \eta/\Omega$. Here, $p_0$ and $f_0$ denote the initial values at $t = 0$, and $\mathrm{Re}[\cdot]$ and $\mathrm{Im}[\cdot]$ denote the real and imaginary parts, respectively. From these expressions, the required derivatives in Eq.~(\ref{eq:rho_deri}) with respect to $t,\omega,\eta$ are obtained as follows. For the derivative with respect to $t$, we have
\begin{eqnarray}
 \partial_t p_t &=& -2\Omega\frac{\partial \Omega}{\partial \eta}\left[\frac{\partial \Omega}{\partial \eta} p_0-\frac{\partial \Omega}{\partial \omega}\mathrm{Re}[f_0]\right]\sin(2\Omega t)-2\Omega\frac{\partial\Omega}{\partial \eta}\mathrm{Im}[f_0]\cos(2\Omega t),\nonumber\\ 
 \mathrm{Re}[\partial_t f_t] &=& 2\Omega\frac{\partial \Omega}{\partial \omega}\left[\frac{\partial \Omega}{\partial \eta} p_0-\frac{\partial \Omega}{\partial \omega}\mathrm{Re}[f_0]\right]\sin(2\Omega t)+2\Omega\frac{\partial\Omega}{\partial \omega}\mathrm{Im}[f_0]\cos(2\Omega t),\nonumber\\
 \mathrm{Im}[\partial_t f_t] &=& -2\Omega\mathrm{Im}[f_0]\sin(2\Omega t)+2\Omega\left[\frac{\partial \Omega}{\partial \eta} p_0-\frac{\partial \Omega}{\partial \omega}\mathrm{Re}[f_0]\right]\cos(2\Omega t).
\end{eqnarray}
For the derivative with respect to $\omega$, we have
\begin{equation}
    \begin{aligned}
    \partial_\omega p_t =& \frac{2}{\Omega}\frac{\partial \Omega}{\partial \eta}\left(2\frac{\partial \Omega}{\partial\omega}\frac{\partial \Omega}{\partial\eta}p_0+\left[\left(\frac{\partial \Omega}{\partial\eta}\right)^2-\left( \frac{\partial \Omega}{\partial \omega}\right)^2\right]\mathrm{Re}[f_0]\right)\sin^2(\Omega t)\\
    &+ \frac{\partial \Omega}{\partial \omega}\frac{\partial \Omega}{\partial\eta}\left(\frac{\mathrm{Im}[f_0]}{\Omega}-2t\left[\frac{\partial \Omega}{\partial\eta}p_0-\frac{\partial \Omega}{\partial\omega}\mathrm{Re}[f_0]\right]\right)\sin(2\Omega t)\\
    &-2t\frac{\partial \Omega}{\partial\omega}\frac{\partial \Omega}{\partial\eta}\mathrm{Im}[f_0]\cos(2\Omega t),
    \end{aligned}
\end{equation}
\begin{equation}  
    \begin{aligned}
     \mathrm{Re}[\partial_\omega f_t]=& \frac{2}{\Omega}\frac{\partial\Omega}{\partial\eta}\left(\left[\left(\frac{\partial\Omega}{\partial\eta}\right)^2-\left(\frac{\partial\Omega}{\partial\omega}\right)^2\right]p_0-2\frac{\partial\Omega}{\partial\omega}\frac{\partial\Omega}{\partial\eta}\mathrm{Re}[f_0]\right)\sin^2(\Omega t) \\
     &+\left(\left(\frac{\partial\Omega}{\partial\eta}\right)^2\frac{\mathrm{Im}[f_0]}{\Omega}+2t\left(\frac{\partial\Omega}{\partial\omega}\right)^2\left[\frac{\partial\Omega}{\partial\eta}p_0-\frac{\partial\Omega}{\partial\omega}\mathrm{Re}[f_0]\right]\right)\sin(2\Omega t) \\
     &+2t\left(\frac{\partial\Omega}{\partial\omega}\right)^2\mathrm{Im}[f_0]\cos(2\Omega t),
    \end{aligned}
\end{equation}
\begin{equation}
    \mathrm{Im}[\partial_\omega f_t] = - \left[2t\frac{\partial\Omega}{\partial \omega}\mathrm{Im}[f_0]+\frac{\partial\Omega}{\partial\omega}\frac{\partial\Omega}{\partial\eta}\frac{p_0}{\Omega}+\left(\frac{\partial\Omega}{\partial\eta}\right)^2\frac{\mathrm{Re}[f_0]}{\Omega}\right]\sin(2\Omega t) + 2t\frac{\partial\Omega}{\partial\omega}\left[\frac{\partial\Omega}{\partial\eta}p_0-\frac{\partial\Omega}{\partial\omega}\mathrm{Re}[f_0]\right]\cos(2\Omega t).
\end{equation}
For the derivative with respect to $\eta$, we have
\begin{equation}
    \begin{aligned}
        \partial_\eta p_t =& -\frac{2}{\Omega}\frac{\partial\Omega}{\partial \omega}\left(2\frac{\partial\Omega}{\partial\omega}\frac{\partial\Omega}{\partial\eta}p_0+\left[\left(\frac{\partial\Omega}{\partial\eta}\right)^2-\left(\frac{\partial\Omega}{\partial\omega}\right)^2\right]\mathrm{Re}[f_0]\right)\sin^2(\Omega t) \\
        &-\left(\left(\frac{\partial\Omega}{\partial\omega}\right)^2\frac{\mathrm{Im}[f_0]}{\Omega}+2t\left(\frac{\partial\Omega}{\partial\eta}\right)^2\left[\frac{\partial\Omega}{\partial\eta}p_0-\frac{\partial\Omega}{\partial\omega}\mathrm{Re}[f_0]\right]\right)\sin(2\Omega t) \\
        &-2t\left(\frac{\partial\Omega}{\partial\eta}\right)^2 \mathrm{Im}[f_0]\cos(2\Omega t),
    \end{aligned}
\end{equation}
\begin{equation}
    \begin{aligned}
        \mathrm{Re}[\partial_\eta f_t] =& -\frac{2}{\Omega} \frac{\partial \Omega}{\partial \omega}\left(\left[\left(\frac{\partial\Omega}{\partial\eta}\right)^2-\left(\frac{\partial\Omega}{\partial\omega}\right)^2\right]p_0-2\frac{\partial\Omega}{\partial\omega}\frac{\partial\Omega}{\partial\eta}\mathrm{Re}[f_0]\right)\sin^2(\Omega t) \\
        &-\frac{\partial\Omega}{\partial\omega}\frac{\partial\Omega}{\partial\eta} \left(\frac{\mathrm{Im}[f_0]}{\Omega}-2t\left[\frac{\partial\Omega}{\partial\eta}p_0-\frac{\partial\Omega}{\partial\omega}\mathrm{Re}[f_0]\right]\right)\sin(2\Omega t) \\
        &+2t\frac{\partial\Omega}{\partial\omega}\frac{\partial\Omega}{\partial\eta}\mathrm{Im}[f_0]\cos(2\Omega t),
    \end{aligned}
\end{equation}
\begin{equation}
    \mathrm{Im}[\partial_\eta f_t] = \left[-2t\frac{\partial\Omega}{\partial \eta}\mathrm{Im}[f_0]+\left(\frac{\partial \Omega}{\partial \omega}\right)^2\frac{p_0}{\Omega}+ \frac{\partial \Omega}{\partial \omega} \frac{\partial \Omega}{\partial \eta}\frac{\mathrm{Re}[f_0]}{\Omega}\right]\sin(2\Omega t) + 2t\frac{\partial\Omega}{\partial\eta}\left[\frac{\partial\Omega}{\partial\eta}p_0-\frac{\partial\Omega}{\partial\omega}\mathrm{Re}[f_0]\right]\cos(2\Omega t).
\end{equation}

Using these explicit expressions, we can directly verify the scaling relation $t\partial_t \rho = \omega \partial_\omega \rho + \eta \partial_\eta \rho$. Following the derivation given in the End Matter of the main text, we then obtain the sum rule,
\begin{equation}
    t^2\mathcal{F}_t~=~\omega^2 \mathcal{F}_\omega +\eta^2\mathcal{F}_\eta + 2(\omega\eta) \mathcal{F}_{\omega \eta},
\end{equation}
where we have used $\mathcal{F}_{\omega \eta}=\mathcal{F}_{\eta\omega}$ to combine cross terms. This is exactly Eq.~(6) of the main text. 

\subsubsection{2.~Tight bound}
Evaluating the tight bound in Eq.~(4) requires computing both the QFI and the response function. According to the compact QFI definition in Eq.~(\ref{eq:QFI_compact}), evaluating QFI reduces to finding the derivative of the system density matrix with respect to the target parameter. As shown in Eq.~(\ref{eq:rho_deri}), this amounts to differentiating the time-dependent density-matrix elements $p_t$ and $f_t$ with respect to that same parameter; these partial derivatives have already been given in the preceding text. With the explicit expressions in Eq.~(\ref{eq:solution_unitary}), we can readily construct the eigenvalues and eigenstates that appear in Eqs.~(\ref{eq:eigenvalue_rho}) and~(\ref{eq:eigenstate_rho}). Substituting these quantities into Eq.~(\ref{eq:QFI_compact}) then directly yields the QFI for each target parameter.

The response function is defined as
\begin{equation}
    \left|\frac{\Delta_x \langle\mathcal{O}\rangle}{\partial_\theta \langle\mathcal{O}\rangle}\right|^2~=~\left|\frac{\mathrm{Tr}[\partial_x\rho \cdot \mathcal{O}]}{\mathrm{Tr}[\partial_\theta\rho \cdot \mathcal{O}]+\mathrm{Tr}[\rho\cdot\partial_\theta\mathcal{O}]}\right|^2,
\end{equation}
where $x\in \{t,\omega,\eta\}$ and $\theta$ is the given target parameter, $\mathcal{O}$ is an arbitrary observable satisfying $\partial_\theta \mathcal{O}=0$. The response functions can be readily evaluated using the derivatives of the state obtained above. With response functions and QFI elements, we can evaluate the tight bound defined in Eq.~(4) of the main text, with the results shown in Fig.~2. 

\subsection{D.~Example II:~Quantum thermometry}
As our second example in the main text, we analyze quantum thermometry which represents a prototypical application of noisy quantum metrology. To obtain unambiguous analytical results, we consider a qubit-based quantum thermometer whose temperature-dependent evolution is governed by the following quantum Lindblad master equation
\begin{equation}
    \partial_t \rho = - i \left[H, \rho \right] + \gamma_{+} \mathcal{D}[J_{+}]\rho + \gamma_{-} \mathcal{D}[J_{-}]\rho + \gamma_{z} \mathcal{D}[J_{z}]\rho,
\end{equation}
where the system Hamiltonian $H=\omega\sigma_z/2$. The three dissipation channels describe thermal excitation $J_{+}=\sigma_{+}$, thermal relaxation $J_{-}=\sigma_{-}$ and pure dephasing $J_{z}=\sigma_z$. The corresponding rates are $\gamma_{+} = \gamma N$, $\gamma_{-} = \gamma (N+1)$ and $\gamma_{z} = \gamma_{0}$, respectively, where $N = \bigl(e^{\beta\omega}-1\bigr)^{-1}$ is the Bose‑Einstein distribution of the thermal bath at inverse temperature $\beta = 1/T$, and $\gamma$ quantifies the overall coupling strength to the bath. In quantum thermometry, the bath temperature $T$ is the target parameter we want to estimate via measuring the dynamics of the thermometer. Hence, in the following, we fix $\theta=T$, while the parameter $x$ in Eq. (4) of the main text ranges over the full parameter set $\{t,\omega,\gamma,\gamma_0,T\}$.

For later convenience, we introduce the combined rates $\gamma_p = \gamma_+ + \gamma_-$ and $\gamma_m = \gamma_+ - \gamma_-$. Based on the parameterization in Eq. (\ref{eq:rho_expression}), the time-dependent solution of the above master equation can be expressed as
\begin{equation}
    \begin{aligned}
        p_t - \frac{\gamma_m}{2 \gamma_p}~=~&\left(p_0 - \frac{\gamma_m}{2 \gamma_p} \right)e^{-\gamma_p t},  \\ 
        f_t~=~&f_0 e^{-i\omega t} e^{-\left(2\gamma_0 + \frac{\gamma_p}{2} \right)t},
    \end{aligned}
\end{equation}
where $p_0$ and $f_0$ are the initial values at $t=0$. In the long‑time limit the system reaches a steady state with $p_{\mathrm{ss}} = \gamma_m/(2\gamma_p)$ and $f_{\mathrm{ss}}=0$, which corresponds to a Gibbsian state at the bath temperature $T$.

\subsubsection{1.~Sum rule for the QFI}
To compute relevant elements of the QFI matrix and verify the sum rule, we first express the SLD operators in the eigenbasis of $\rho$. Noting Eq. (\ref{eq:Fx_simple}), it suffices to give the matrices $U^{\dagger}L_xU$ with $x\in\{t,\omega,\gamma,\gamma_0,T\}$:
\begin{equation}
    \begin{aligned}\label{eq:uu2}
    U^\dagger L_{\gamma} U~=~&a_{\gamma}\mathbb{I}+b_{\gamma}\sigma_x+d_{\gamma}\sigma_z, \\
    U^\dagger L_{\gamma_0} U~=~& a_{\gamma_0}\mathbb{I}+b_{\gamma_0}\sigma_x+d_{\gamma_0}\sigma_z,\\
    U^\dagger L_{T} U~=~&a_{T}\mathbb{I}+b_{T}\sigma_x+d_{T}\sigma_z, \\
    U^\dagger L_{\omega} U~=~& a_{\omega}\mathbb{I}+b_{\omega}\sigma_x+c_{\omega}\sigma_y+d_{\omega}\sigma_z, \\
    U^\dagger L_{t} U~=~& a_{t}\mathbb{I}+b_{t}\sigma_x+c_{t}\sigma_y+d_{t}\sigma_z,
    \end{aligned}
\end{equation}
where all coefficients are real functions of $p_t$, $|f_t|$, the eigenvalues $\mu_\pm$ of state, and the parameters $\{t,\omega,\gamma,\gamma_0,T\}$. Their explicit forms are listed below.
\begin{itemize}
    \item [(i)] For the damping parameters $\gamma$ and $\gamma_0$, the coefficients read
    \begin{equation}
    a_{\gamma}~=~-\frac{t\gamma_p}{\gamma_m}\frac{\left(p_t-\frac{\gamma_m}{2\gamma_p} \right)p_t + \frac{1}{2}|f_t|^2}{\mu_+\mu_-},~~
    b_{\gamma}~=~\frac{t\gamma_p}{\gamma_m}\frac{ \left( p_t-\frac{\gamma_m}{\gamma_p}\right)|f_t|}{\sqrt{p_t^2+|f_t|^2}},~~
    d_{\gamma}~=~ -\frac{a_\gamma}{2\sqrt{p_t^2+|f_t|^2}},
    \end{equation}
    \begin{equation}
        a_{\gamma_0}~=~\frac{2t|f_t|^2}{\mu_+\mu_-},~~
        b_{\gamma_0}~=~\frac{4tp_t|f_t|}{\sqrt{p_t^2+|f_t|^2}},~~
        d_{\gamma_0}~=~-\frac{a_{\gamma_0}}{2\sqrt{p_t^2+|f_t|^2}}.
    \end{equation}
    \item [(ii)] For the temperature $T$ and the frequency $\omega$, the coefficients are more involved but share a similar structure,
    \begin{equation}
    \begin{aligned}
    a_{T}&~=~\frac{\omega  \gamma_+\gamma_-}{T^2\gamma_p^2\mu_+\mu_-}\left[ p_t \left(e^{-\gamma_p t}+\gamma_p t-1 \right) - \frac{\gamma_p^2 t}{\gamma_m}\left(2p_t^2+ |f_t|^2\right)\right], \\
    b_{T}&~=~-2\frac{\omega  \gamma_+\gamma_-|f_t|}{T^2\gamma_p^2\sqrt{p_t^2+|f_t|^2}} \left(e^{-\gamma_p t}+\gamma_p t-1-\frac{\gamma_p^2 t}{\gamma_m} p_t\right), \\
    d_{T}&~=~ -\frac{a_{T}}{2\sqrt{p_t^2+|f_t|^2}},
    \end{aligned}
    \end{equation}
    \begin{equation}
        \begin{aligned}
        a_{\omega}~=~&-\frac{  \gamma_+\gamma_-}{T\gamma_p^2\mu_+\mu_-} \left[ p_t \left(e^{-\gamma_p t}+\gamma_p t-1 \right) - \frac{\gamma_p^2 t}{\gamma_m}\left(2p_t^2+ |f_t|^2\right)\right],~~\\
        b_{\omega}~=~&2\frac{ \gamma_+\gamma_-|f_t|}{ T\gamma_p^2\sqrt{p_t^2+|f_t|^2}} \left(e^{-\gamma_p t}+\gamma_p t-1-\frac{\gamma_p^2 t}{\gamma_m} p_t\right),~~\\
        c_{\omega}~=~&-2t|f_t|,~~
        d_{\omega}~=~-\frac{a_{\omega}}{2\sqrt{p_t^2+|f_t|^2}}.
        \end{aligned}
    \end{equation}
    \item [(iii)] For the time $t$, the coefficients are
    \begin{equation}
    \begin{aligned}
    a_t~=~&-\frac{\left(2\gamma_0+\frac{\gamma_p}{2} \right)|f_t|^2+ \left( p_t - \frac{\gamma_m}{2\gamma_p} \right)\gamma_p p_t}{2\mu_+ \mu_-},\\
    b_t~=~&-\frac{\left[ \left(4\gamma_0 - \gamma_p \right)p_t+\gamma_m \right]|f_t|}{\sqrt{p_t^2+|f_t|^2}}, \\
    c_t~=~&-2\omega |f_t|,~~
    d_t~=~-\frac{a_t}{2\sqrt{p_t^2+|f_t|^2}}.
    \end{aligned}
    \end{equation}
\end{itemize}
Interestingly, Eq. (\ref{eq:uu2}) reveals several linear relations among the SLD matrices $U^{\dagger}L_xU$. First, we have
\begin{equation}\label{eq:QT_SLD_relation_1}
    t\left(U^\dagger L_t U\right)~=~\gamma \left(U^\dagger L_{\gamma} U\right) +\gamma_0 \left(U^\dagger L_{\gamma_0} U\right) + \omega c_{\omega}\sigma_y.
\end{equation}
Second, from the expressions for $U^\dagger L_{\omega} U$ and $U^\dagger L_{T} U$ we observe
\begin{equation}\label{eq:QT_SLD_relation_2}
    \omega c_{\omega}\sigma_y~=~\omega \left(U^\dagger L_{\omega} U\right)+T \left( U^\dagger L_{T} U\right).
\end{equation}
Combining Eqs.~(\ref{eq:QT_SLD_relation_1}) and (\ref{eq:QT_SLD_relation_2}) yields the linear identity
\begin{equation}\label{eq:QT_SLD_relation}
    t\left(U^\dagger L_t U\right)~=~\gamma \left(U^\dagger L_{\gamma} U\right) +\gamma_0 \left(U^\dagger L_{\gamma_0} U\right) + \omega \left(U^\dagger L_{\omega} U\right)+T \left( U^\dagger L_{T} U\right),
\end{equation}
which mirrors the derivative relation 
\begin{equation}
    t\partial_t \rho~=~\omega \partial_\omega \rho + T\partial_T\rho +\gamma \partial_\gamma \rho +\gamma_0 \partial_{\gamma_0} \rho.
\end{equation}

Inserting Eq. (\ref{eq:uu2}) into Eq. (\ref{eq:Fx_simple}) gives the elements of the QFI matrix. The diagonal elements (the QFI about each parameter) are
\begin{equation}
    \mathcal{F}_{\gamma}~=~b_{\gamma}^2+d_{\gamma}^2-a_{\gamma}^2,~~
    \mathcal{F}_{\gamma_0}~=~b_{\gamma_0}^2+d_{\gamma_0}^2-a_{\gamma_0}^2,~~
    \mathcal{F}_{T}~=~b_T^2+d_T^2-a_T^2,~~
    \mathcal{F}_{\omega}~=~b_\omega^2+c_\omega^2+d_\omega^2-a_\omega^2,
\end{equation}
while $\mathcal{F}_t$ will be given below. The non-vanishing off-diagonal elements read
\begin{equation}
\mathcal{F}_{\gamma\gamma_0}~=~b_{\gamma}b_{\gamma_0}+d_{\gamma}d_{\gamma_0}-a_{\gamma}a_{\gamma_0},~~
    \mathcal{F}_{\gamma T}~=~b_{\gamma}b_{T}+d_{\gamma}d_{T}-a_{\gamma}a_{T},~~\mathcal{F}_{\gamma\omega}~=~b_{\gamma}b_{\omega}+d_{\gamma}d_{\omega}-a_{\gamma}a_{\omega},~~
\end{equation}
\begin{equation}
    \mathcal{F}_{\gamma_0 T}~=~b_{\gamma_0}b_{T}+d_{\gamma_0}d_{T}-a_{\gamma_0}a_{T},~~
    \mathcal{F}_{\gamma_0\omega}~=~b_{\gamma_0}b_{\omega}+d_{\gamma_0}d_{\omega}-a_{\gamma_0}a_{\omega},~~
    \mathcal{F}_{\omega T}~=~b_{\omega}b_{T}+d_{\omega}d_{T}-a_{\omega}a_{T}.
\end{equation}
From Eq.~(\ref{eq:QT_SLD_relation_1}), we can obtain a compact relation
\begin{equation}\label{eq:QFI_1}
    t^2 \mathcal{F}_t~=~\gamma^2 \mathcal{F}_{\gamma}+\gamma_0^2 \mathcal{F}_{\gamma_0}+2\gamma \gamma_0 \mathcal{F}_{\gamma \gamma_0} +\omega^2 c_\omega^2.
\end{equation}
The term $\omega^{2}c_{\omega}^{2}$ can itself be expressed through the QFI about temperature and frequency by noting Eq.~(\ref{eq:QT_SLD_relation_2}),
\begin{equation}\label{eq:QFI_2}
    \omega^2 c_\omega^2~=~T^2 \mathcal{F}_{T}+\omega^2 \mathcal{F}_{\omega}+2T\omega \mathcal{F}_{T \omega}.
\end{equation}
Substituting Eq. (\ref{eq:QFI_2}) into Eq. (\ref{eq:QFI_1}) yields the sum rule for the QFI matrix
\begin{equation}\label{eq:QFI}
    t^2 \mathcal{F}_t~=~\gamma^2 \mathcal{F}_{\gamma}+\gamma_0^2 \mathcal{F}_{\gamma_0}+ T^2 \mathcal{F}_{T}+\omega^2 \mathcal{F}_{\omega}+2\gamma \gamma_0 \mathcal{F}_{\gamma \gamma_0}+2T\omega \mathcal{F}_{T \omega},
\end{equation}
which is Eq. (8) of the main text and confirms the general sum rule Eq. (3) of the main text.

It is instructive to examine the signs of the two off-diagonal terms appearing in the above sum rule. For $\mathcal{F}_{\gamma\gamma_0}$, we find
\begin{equation}\label{eq:QFI_gamma_and_gamma0}
    \mathcal{F}_{\gamma\gamma_0}~=~-\frac{4\gamma_p t^2 |f_t|^2}{\gamma_m \mu_+\mu_-}  \mathcal{A}\left(p_t\right),
\end{equation}
where $\mathcal{A}\left(p_t\right)=p_t^2-2p_{ss}p_t+1/4$ with $p_t\in [-1/2, 1/2]$. The quadratic $\mathcal{A}\left(p_t\right)$ attains its minimum at the steady‑state value $p_t = p_{ss}$; because $\mathcal{A}_{ss}\left(p_t\right)=\mathcal{A}\left(p_{ss}\right)=1/4-p_{ss}^2>0$, we conclude $\mathcal{F}_{\gamma\gamma_0}>0$. As for $\mathcal{F}_{T\omega}$, we obtain the simple relation
\begin{equation}\label{eq:QFI_T_and_omega}
    \mathcal{F}_{T\omega}~=~-\frac{T}{\omega}\mathcal{F}_T~<~0.
\end{equation}

\subsubsection{2.~Tight bound for observable \texorpdfstring{$H$}{Hamiltonian}}
For quantum thermometry, the system Hamiltonian happens to be the optimal observable that saturates the corresponding QCRB $\mathcal{F}_T^{-1}$ at thermal equilibrium. We therefore fix the observable $\mathcal{O}=H$ in Eq. (4) of the main text and examine whether the QCRB $\mathcal{F}_T^{-1}$ remains the tight bound at finite times. Because $\langle H\rangle = \omega p_t$ and $H$ depends on the parameter $\omega$, we must use the modified derivative $\Delta_x\langle H\rangle\equiv\partial_x\langle H\rangle -\langle \partial_x H\rangle$ introduced in the main text. The relevant derivatives are
\begin{equation}
    \begin{aligned}
    \Delta_{\gamma_0} \langle H\rangle~=~& \partial_{\gamma_0} \langle H\rangle~=~ 0,~~\\
    \Delta_t \langle H\rangle~=~& \partial_{t} \langle H\rangle~=~-\omega\gamma_p \left(p_t -p_{ss}\right),~~ \\
    \Delta_\gamma \langle H\rangle~=~& \partial_{\gamma} \langle H\rangle~=~\omega  \frac{\gamma_p t}{\gamma_m} \left(p_t -p_{ss}\right),~~\\
    \Delta_\omega \langle H\rangle~=~& -\frac{\omega\gamma_+\gamma_-}{\gamma_m^2T}\left[ \frac{\gamma_m^2}{\gamma_p^2} \left( 1-e^{-\gamma_p t}\right)+2\gamma_m t\left(p_t -p_{ss}\right)\right],~~\\
    \Delta_{T} \langle H\rangle~=~& \partial_{T} \langle H\rangle~=~ \frac{\omega^2\gamma_+\gamma_-}{\gamma_m^2T^2}\left[ \frac{\gamma_m^2}{\gamma_p^2} \left( 1-e^{-\gamma_p t}\right)+2\gamma_m t\left(p_t -p_{ss}\right)\right].
    \end{aligned}
\end{equation}
From the above derivatives we observe that $t\Delta_t \langle H\rangle=\gamma \Delta_\gamma \langle H\rangle$ and $\omega\Delta_\omega \langle H\rangle+T \Delta_T \langle H\rangle=0$.

Applying Eq. (4) of the main text, we know that identifying the tightest bound for estimating the temperature $T$ amounts to comparing five candidate bounds,
\begin{equation}\label{eq:BBB_2_all}
     \max_{x \in \{\gamma_0, t, \gamma, \omega, T\}} \left
    \{\frac{1}{\mathcal{F}_x}\left| \frac{\Delta_x \langle H\rangle}{\partial_T \langle H\rangle}\right|^2\right\}~=~\frac{1}{|\partial_T \langle H\rangle|^2} \max \left\{ 0,~~
    \frac{|\Delta_t \langle H\rangle|^2}{\mathcal{F}_t},~~
    \frac{|\Delta_\gamma \langle H\rangle|^2}{\mathcal{F}_\gamma},~~
    \frac{|\Delta_\omega \langle H\rangle|^2}{\mathcal{F}_\omega},~~
    \frac{|\Delta_T \langle H\rangle|^2}{\mathcal{F}_T} \right\}.
\end{equation}
Using the QFI relations Eqs.~(\ref{eq:QFI_1}) and~(\ref{eq:QFI_gamma_and_gamma0}), we obtain
\begin{equation}
     \max \left\{
    \frac{|\Delta_t \langle H\rangle|^2}{\mathcal{F}_t},~~
    \frac{|\Delta_\gamma \langle H\rangle|^2}{\mathcal{F}_\gamma}
    \right\}~=~\max \left\{
    \frac{\gamma^2|\Delta_\gamma \langle H\rangle|^2}{\gamma^2 \mathcal{F}_{\gamma}+\gamma_0^2 \mathcal{F}_{\gamma_0}+2\gamma \gamma_0 \mathcal{F}_{\gamma \gamma_0} +\omega^2 c_\omega^2},~~
    \frac{|\Delta_\gamma \langle H\rangle|^2}{\mathcal{F}_\gamma}
    \right\}~=~\frac{|\Delta_\gamma \langle H\rangle|^2}{\mathcal{F}_\gamma}.
\end{equation}
Similarly, from Eqs.~(\ref{eq:QFI_2}) and~(\ref{eq:QFI_T_and_omega}), we have
\begin{equation}
    \max \left\{
    \frac{|\Delta_\omega \langle H\rangle|^2}{\mathcal{F}_\omega},~~
    \frac{|\Delta_T \langle H\rangle|^2}{\mathcal{F}_T}
    \right\}~=~\max \left\{
    \frac{T^2|\Delta_T \langle H\rangle|^2}{T^2\mathcal{F}_T+\omega^2c_\omega^2},~~
    \frac{|\Delta_T \langle H\rangle|^2}{\mathcal{F}_T}
    \right\}~=~\frac{|\Delta_T \langle H\rangle|^2}{\mathcal{F}_T}.
\end{equation}
Thus Eq. (\ref{eq:BBB_2_all}) reduces to a comparison of the bounds for $x=\gamma$ and $x=T$,
\begin{equation}\label{eq:BBB_2_last}
    \max_{x \in \{\gamma_0, t, \gamma, \omega, T\}} \left
    \{\frac{1}{\mathcal{F}_x}\left| \frac{\Delta_x \langle H\rangle}{\partial_T \langle H\rangle}\right|^2\right\}~=~\frac{1}{|\partial_T \langle H\rangle|^2} \max \left\{ 
    \frac{|\Delta_\gamma \langle H\rangle|^2}{\mathcal{F}_\gamma},~~
    \frac{|\Delta_T \langle H\rangle|^2}{\mathcal{F}_T}  
    \right\}.
\end{equation}
Since $\partial_T \langle H\rangle=\Delta_T\langle H\rangle$, Eq. (\ref{eq:BBB_2_last}) can be written compactly as
\begin{equation}\label{eq:BBB_2}
    \max_{x \in \{\gamma_0, t, \gamma, \omega, T\}} \left
    \{\frac{1}{\mathcal{F}_x}\left| \frac{\Delta_x \langle H\rangle}{\partial_T \langle H\rangle}\right|^2\right\}~=~\frac{1}{\mathcal{F}_T + \mathcal{R}},
\end{equation}
where we have defined
\begin{equation}
    \mathcal{R}~=~
    \min \left\{ 
    \mathcal{F}_\gamma \left|\frac{\Delta_T \langle H\rangle}{\Delta_\gamma \langle H\rangle}\right|^2 - \mathcal{F}_T,~~
    0
    \right\},
\end{equation}
which takes the minimum over the two terms in the bracket. From Eq. (\ref{eq:BBB_2}), it is clear that the QCRB $\mathcal{F}_T^{-1}$ ceases to be the tightest bound at finite times whenever $\mathcal{R}\neq 0$, i.e., when $\mathcal{F}_\gamma \left|\frac{\Delta_T \langle H\rangle}{\Delta_\gamma \langle H\rangle}\right|^2 -\mathcal{F}_T<0$. Because the value of $\mathcal{R}$ depends on initial states, the tightness of the QCRB during the nonequilibrium evolution is itself initial‑state dependent.

The explicit expressions needed to evaluate the sign of $\mathcal{R}$ are:
\begin{itemize}
    \item [(i)] \begin{equation}
    \mathcal{F}_\gamma \left|\frac{\Delta_T \langle H\rangle}{\Delta_\gamma \langle H\rangle}\right|^2~=~\frac{4\omega^2 t^2 \gamma_+^2\gamma_-^2}{ T^4\gamma_p^2p_{ss}^2(1-4\mu_+\mu_-)}  \left[\left(2p_{ss}\mathcal{H}-p_t+ \frac{p_tp_{ss}(1-\mathcal{H})}{p_t-p_{ss}}  \right)^2 |f_t|^2+
    \frac{\left[p_t(p_{ss}\mathcal{H}-p_t)-\frac{p_t-p_{ss}\mathcal{H}}{p_t-p_{ss}}\frac{|f_t|^2}{2}\right]^2}{\mu_+\mu_-}\right],
\end{equation}
where $\mathcal{H}= (e^{-\gamma_p t}+\gamma_p t -1)/(\gamma_p t)$.
    \item [(ii)] \begin{equation}\label{eq:m3_QFI_T}
    \mathcal{F}_T ~=~\frac{4\omega^2 t^2  \gamma_+^2\gamma_-^2}{ T^4\gamma_p^2p_{ss}^2(1-4\mu_+\mu_-)}  \left[(2p_{ss}\mathcal{H}-p_t)^2|f_t|^2+\frac{\left[ p_t(p_{ss}\mathcal{H}-p_t)-\frac{1}{2}|f_t|^2\right]^2}{\mu_+\mu_-}\right].
\end{equation}
\end{itemize}
Their difference simplifies to
\begin{equation}\label{eq:eee}
    \mathbb{E}~=~\mathcal{F}_\gamma \left|\frac{\Delta_T \langle H\rangle}{\Delta_\gamma \langle H\rangle}\right|^2-\mathcal{F}_T ~=~\frac{\omega^2 t^2  \gamma_+^2\gamma_-^2}{T^4\gamma_p^2p_{ss}^2\mu_+\mu_-} 
    \frac{ p_{ss} \left(1-\mathcal{H} \right) |f_t|^2}{p_t-p_{ss}} 
    \left[ 4p_t(p_t -p_{ss}\mathcal{H}) +\left(\frac{1}{4}-p_t^2\right) \frac{2p_t-p_{ss}(\mathcal{H}+1)}{p_t-p_{ss}}\right].
\end{equation}
Eq.~(\ref{eq:eee}) is used in the main text to identify the parameter regimes where $\mathcal{R}<0$ or $\mathcal{R}=0$; the resulting dependence on the initial state is displayed in Fig.~3 of the main text.

\subsection{E.~Example III: Two-qubit model}
This subsection presents a numerical demonstration of Eq.~(4) of the main text using a two-qubit model that is not analytically solvable. We first describe a general numerical strategy for computing the QFI $\mathcal{F}_x$ and the response functions $\Delta_x\langle\mathcal{O}\rangle$ for each parameter $x$, and then numerically identify the tight precision bound for the two-qubit model by evaluating all candidate bounds.

\subsubsection{1.~Two-qubit model and metrological setting}
We consider an open two-qubit system that generalizes the single-qubit spontaneous emission model analyzed in Example II. The system Hamiltonian reads
\begin{equation}
    H(\omega, \lambda)~=~\omega \left(\sigma_z^{(1)} + \sigma_z^{(2)}\right) + \lambda \, \sigma_x^{(1)}\sigma_x^{(2)},
\end{equation}
where $\omega$ is the local energy gap which is assumed to be identical for both qubits, $\lambda$ is the spin-spin coupling strength, and we use the short-hand notations $\sigma_z^{(1)} \equiv \sigma_z^{(1)} \otimes \mathbb{I}$, $\sigma_z^{(2)} \equiv \mathbb{I} \otimes \sigma_z^{(2)}$, and $\sigma_x^{(1)}\sigma_x^{(2)} \equiv \sigma_x^{(1)} \otimes \sigma_x^{(2)}$ with superscript $(j)$ labeling the $j$-th qubit. The dissipative evolution of the reduced system state is governed by the following quantum Lindblad master equation
\begin{equation}\label{eq:master_two}
    \partial_t \rho~=~-i[H(\omega, \lambda), \rho] + \gamma \left(\mathcal{D}[\sigma_-^{(1)}]\rho + \mathcal{D}[\sigma_-^{(2)}]\rho\right),
\end{equation}
where $\gamma$ is the spontaneous emission rate and the dissipator $\mathcal{D}[J]\rho = J\rho J^\dagger - \frac{1}{2}\{J^\dagger J, \rho\}$. For this model, the complete parameter set is $\boldsymbol{\Theta} = \{t, \omega, \lambda, \gamma\}$. We remark that this model lacks an analytical treatment.

We aim to estimate the coupling strength $\lambda$, i.e., the target parameter $\theta = \lambda$. In principle, the numerical strategy outlined below applies to an arbitrary observable. Here, as an example, we choose the local energy observable $\mathcal{O} = H^{(1)} = \omega \sigma_z^{(1)}$ and employ Eq.~(4) of the main text to determine the tight precision bound. For demonstration purposes, we take the initial product state $\rho(0) = \ket{gg}\bra{gg}$, where $\ket{gg} \equiv \ket{g}^{(1)} \otimes \ket{g}^{(2)}$ and $|g\rangle^{(j)}$ is the ground state of $\sigma_z^{(j)}$.

\subsubsection{2.~General numerical strategies for evaluating QFI and response functions of observables}
We here present an efficient numerical method for evaluating the QFI and response functions involved in Eq. (4) of the main text. The approach requires only two time-evolved reduced system states, $\rho_{x-\delta x}$ and $\rho_{x+\delta x}$, from which  the dynamics of the QFI $\mathcal{F}_x$ and the response function $\Delta_x\langle\mathcal{O}\rangle$ for a given parameter $x$ can be extracted simultaneously. To get dynamics of $\rho_{x-\delta x}$ and $\rho_{x+\delta x}$, we integrate Eq. (\ref{eq:master_two}) with the parameter $x$ set to $x-\delta x$ and $x+\delta x$, respectively, while keeping all other parameters identical in the two sets of simulations. In simulations, we will vary the value of $\delta x$ until numerical convergence is reached. We emphasize that the numerical scheme described below applies to arbitrary systems and initial states.
\\
\begin{center}
{\it 2.1~Evaluating the quantum Fisher information}
\end{center}
To compute the QFI numerically, we adopt the fidelity-based definition that connects the QFI with the Uhlmann‑Jozsa fidelity $F_U(\rho_x, \rho_{x+\delta x})\equiv\left(\mathrm{Tr}\left[\sqrt{\sqrt{\rho_x}\rho_{x+\delta x}\sqrt{\rho_x}}\right]\right)^2$ between quantum states $\rho_{x}$ and $\rho_{x+\delta x}$ via~\cite{Sidhu.20.AVSQ}
\begin{equation}
    \mathcal{F}_{x}~=~\frac{8}{\left( \delta x\right)^2}\left[ 1 - \sqrt{F_U(\rho_x, \rho_{x+\delta x})}\right].
\end{equation}
For better numerical stability, we rewrite the above definition equivalently as
\begin{equation}\label{eq:fx_num}
    \mathcal{F}_{x}~=~\frac{2}{\left( \delta x\right)^2}\left[ 1 - \sqrt{F_U(\rho_{x-\delta x}, \rho_{x+\delta x})}\right]~=~\frac{2}{\left( \delta x\right)^2}\frac{1 - F_U(\rho_{x-\delta x}, \rho_{x+\delta x})}{1 + \sqrt{F_U(\rho_{x-\delta x}, \rho_{x+\delta x})}}.
\end{equation}
In all simulations, we evaluate the QFI $\mathcal{F}_{x}$ using Eq. (\ref{eq:fx_num}). The Uhlmann‑Jozsa fidelity $F_U$ itself is computed with a recently developed, numerically efficient technique~\cite{Hauru.23.PRA}.

\begin{figure}[h]
 \centering
\includegraphics[width=0.5\columnwidth]{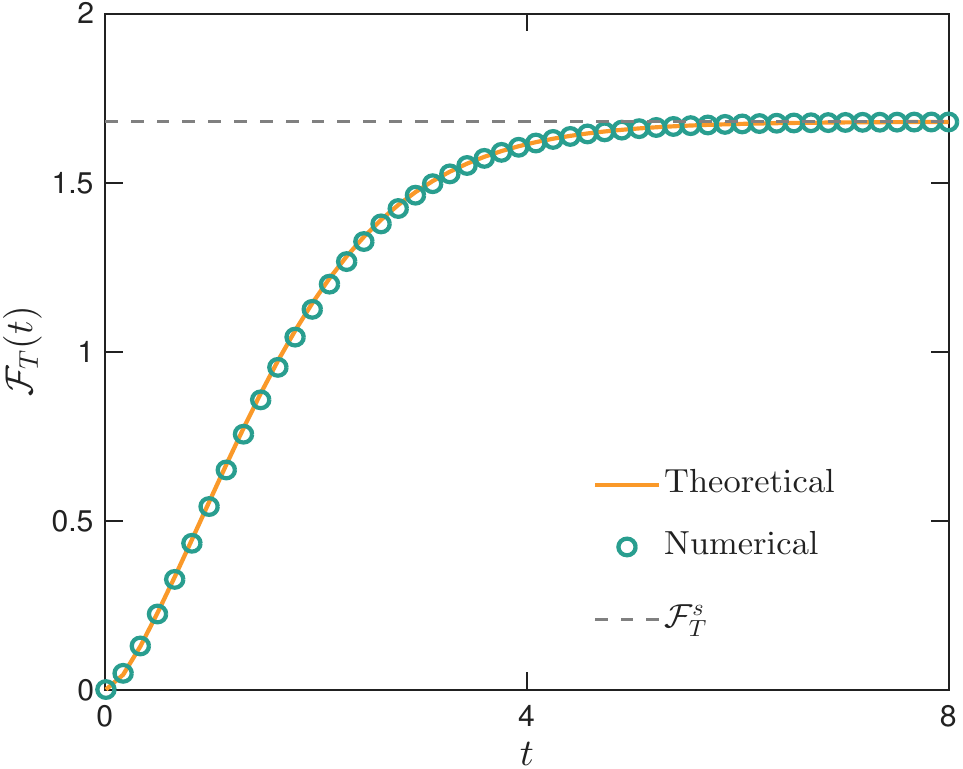} 
 \caption{Dynamics of QFI $\mathcal{F}_T(t)$ for the quantum thermometry model analyzed in Sec. II. Solid line: theoretical result from Eq.~\eqref{eq:m3_QFI_T}; teal circles: numerical evaluation using Eq.~\eqref{eq:fx_num}. The simulation uses dephasing strength $\gamma_0 = 0.2$ and starts from a
coherent initial state $\rho(0)=0.5\mathbb{I}+0.4\sigma_x-0.2\sigma_z$ with $\mathbb{I}$ the identity
matrix. The black dashed line marks
the steady state QFI $\mathcal{F}_T^s=\gamma_+\gamma_-\gamma_p^{-2}\omega^2$ reached as $t \to \infty$. Other parameters are $\omega=1$, $T=0.5$, $\gamma=1$ and $\delta T = 0.001$.}
\protect\label{fig:m3_QFI_T}
\end{figure}

As a benchmark, we first apply this strategy to the quantum thermometry model studied in Example II of the main text, for which an analytical expression for $\mathcal{F}_T$ is available [Eq.~\eqref{eq:m3_QFI_T}]. The numerically converged results are displayed in Fig.~\ref{fig:m3_QFI_T}. The figure shows excellent agreement between the theoretical prediction and the numerical evaluation, confirming the reliability of the definition Eq.~\eqref{eq:fx_num}.

\begin{figure}[h]
 \centering
\includegraphics[width=0.75\columnwidth]{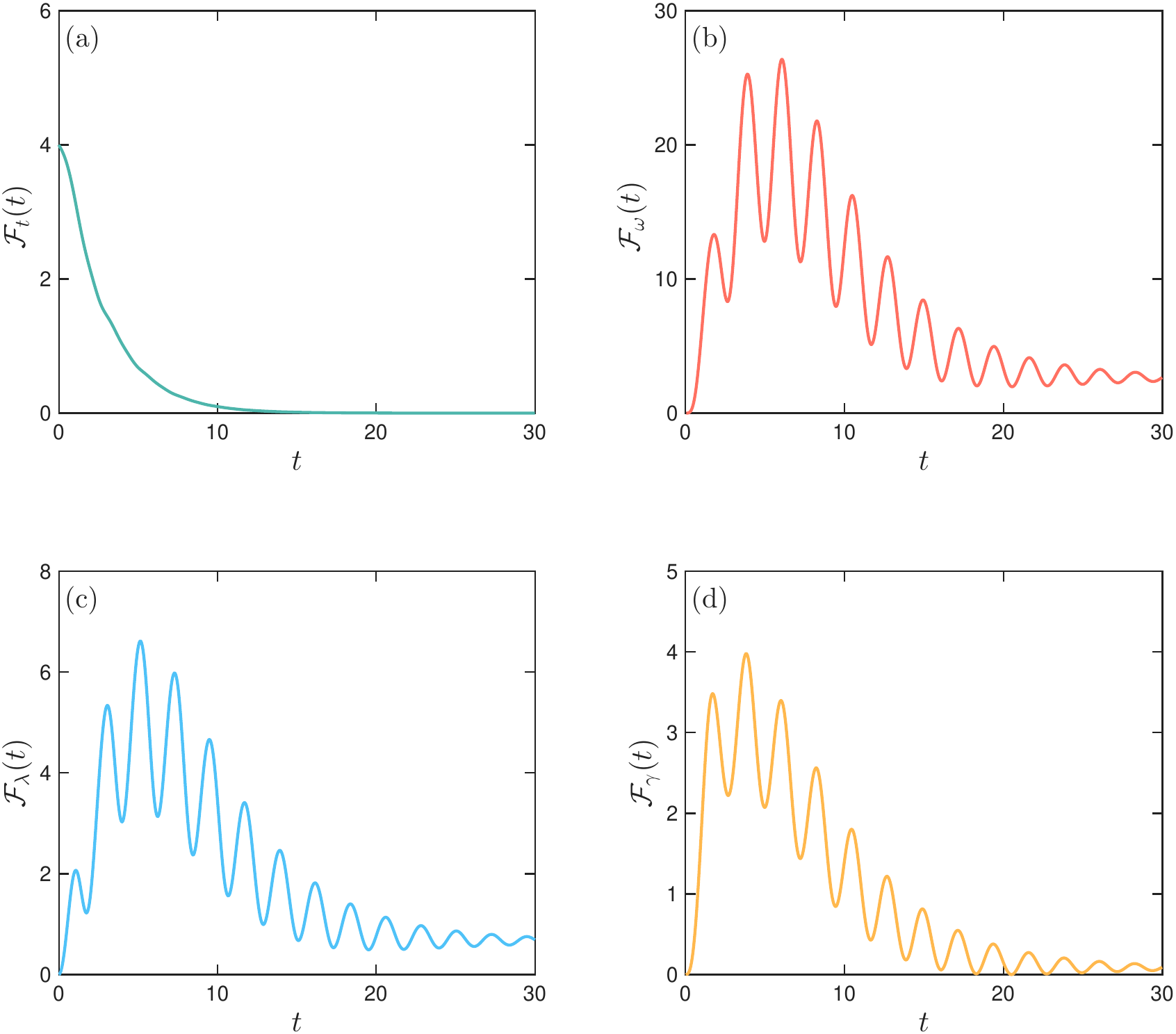} 
 \caption{Dynamics of the QFI for the two-qubit model: (a) $\mathcal{F}_t(t)$, (b) $\mathcal{F}_\omega(t)$, (c) $\mathcal{F}_\lambda(t)$ and (d) $\mathcal{F}_\gamma(t)$. The initial state is $\rho(0)=\ket{gg}\bra{gg}$. Parameters are $\omega=0.5$, $\lambda=1$, $\gamma=0.2$, $\delta \omega =\delta \lambda =\delta \gamma = 0.001$ and $\delta t = 0.0001$.}
\protect\label{fig:QFIs}
\end{figure}

We now use the method to evaluate the QFI $\mathcal{F}_x$ with $x\in\{t,\omega,\lambda,\gamma\}$ for the two-qubit model using Eq. (\ref{eq:master_two}). A set of converged numerical results is showed in Fig. \ref{fig:QFIs}. From the figure, we observe that the QFI for different parameters exhibit distinct dynamical features. $\mathcal{F}_t$ in Fig. \ref{fig:QFIs} (a) decays almost monotonically from its initial maximum, asymptotically approaching zero--a signature of irreversible information loss to the dissipative environment. By contrast, the QFIs associated with the Hamiltonian parameters $\omega$ and $\lambda$, as well as with the spontaneous emission rate $\gamma$ [Fig. \ref{fig:QFIs} (b), (c) and (d), respectively], all display non-monotonic oscillations. Notably, their local maxima and minima do not coincide in time.
\\
\begin{center}
{\it 2.2~Evaluating the response functions of observables}
\end{center}

The response functions can be evaluated utilizing the same set of dynamical states, $\rho_{x\pm \delta x}$, that are already obtained when computing the QFI, thereby eliminating redundant numerical effort. To illustrate the formalism, we take $\mathcal{O}=H^{(1)}$ as an example. We note that $\Delta_x \langle H^{(1)} \rangle \equiv \partial_x \langle H^{(1)} \rangle- \langle \partial_xH^{(1)} \rangle= \mathrm{Tr}[(\partial_x\rho)H^{(1)}]$, which can be computed via the finite-difference expression
\begin{equation}
    \Delta_x \langle H^{(1)}  \rangle = \frac{\mathrm{Tr}\left[\left(\rho_{x+\delta x} - \rho_{x-\delta x}\right)\cdot H^{(1)} \right]}{2\delta x}.
\end{equation}
Furthermore, we remark that $\Delta_x \langle H^{(1)} \rangle$ satisfies the scaling relation
\begin{equation}\label{eq:scaling_H1}
    t\Delta_t\langle H^{(1)}\rangle~=~\sum_{\theta\in \{ \omega,\lambda,\gamma\}} \theta\Delta_\theta\langle H^{(1)}\rangle,
\end{equation}
obtained by multiplying both sides of Eq.~(A4) of the end matter by the observable $H^{(1)}$ and taking the trace. Eq. (\ref{eq:scaling_H1}) provides a direct benchmark for verifying the numerical results for $\Delta_x \langle H^{(1)} \rangle$.

\begin{figure}[h]
 \centering
\includegraphics[width=0.90\columnwidth]{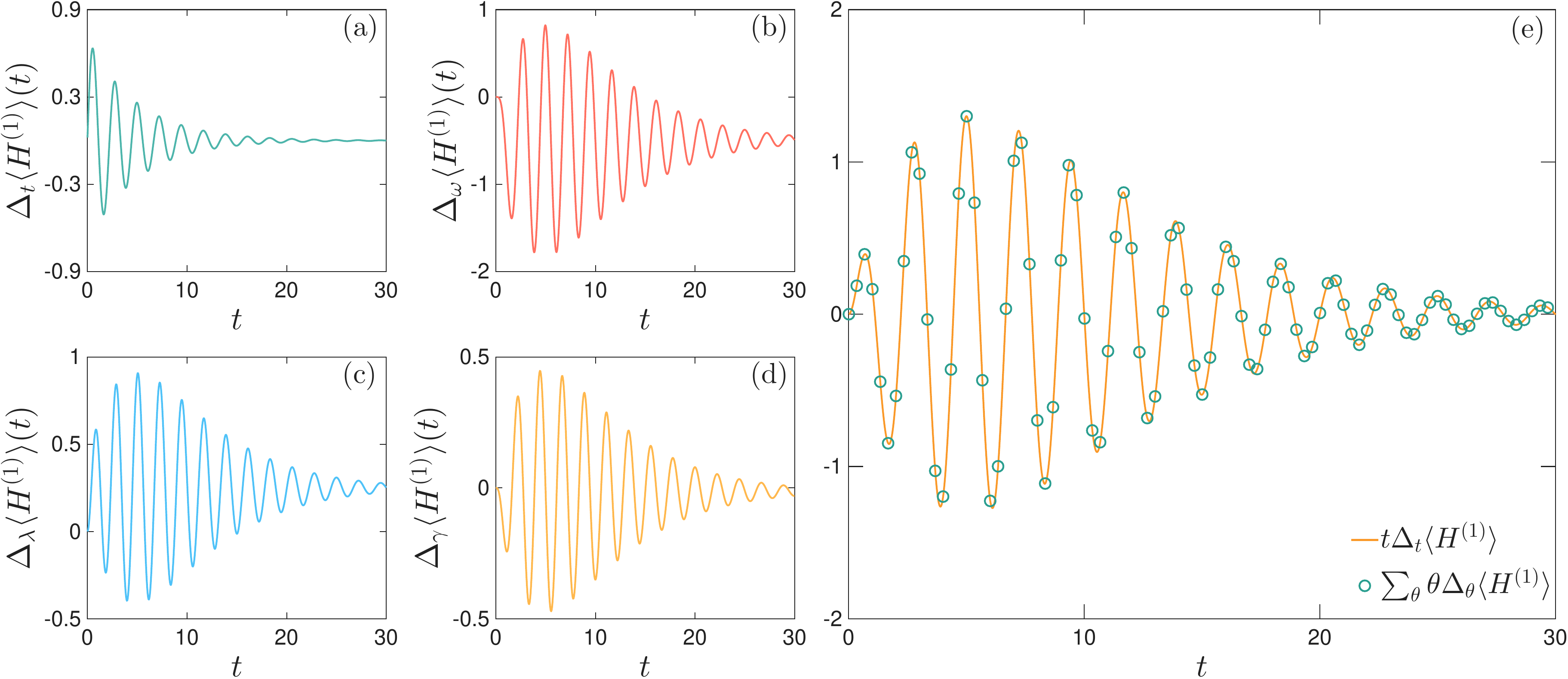} 
 \caption{Dynamics of response functions: (a) $\Delta_t\langle H^{(1)}\rangle(t)$, (b) $\Delta_\omega\langle H^{(1)}\rangle(t)$, (c) $\Delta_\lambda\langle H^{(1)}\rangle(t)$ and (d) $\Delta_\gamma\langle H^{(1)}\rangle(t)$. (e) Verification of the scaling relation: the solid line represents $t\Delta_t\langle H^{(1)}\rangle$, and the circle denotes $\sum_\theta \theta\Delta_\theta\langle H^{(1)}\rangle$ with $\theta \in \{\omega, \lambda, \gamma\}$. The initial state is $\rho(0)=\ket{gg}\bra{gg}$. Parameters are $\omega=0.5$, $\lambda=1$, $\gamma=0.2$, $\delta \omega =\delta \lambda =\delta \gamma = 0.001$ and $\delta t = 0.0001$.}
\protect\label{fig:Scaling}
\end{figure}

A set of numerically converged results for the two-qubit model is summarized in Fig.~\ref{fig:Scaling}. Panels (a)--(d) display the dynamics of $\Delta_x\langle H^{(1)}\rangle$ for $x = t, \omega, \lambda, \gamma$, respectively, all of which exhibit damped oscillations. Panel (e) tests the relation in Eq. (\ref{eq:scaling_H1}), showing excellent agreement between $t\Delta_t\langle H^{(1)}\rangle$ (solid line) and the sum $\sum_{\theta}\theta\Delta_\theta\langle H^{(1)}\rangle$ ( circle) over $\theta \in \{\omega, \lambda, \gamma\}$ during the evolution, thereby validating the numerical results in panels (a)--(d).

Consequently, the ratios appearing in the precision bounds in Eq. (4) of the main text can be evaluated as
\begin{equation}
    \frac{\Delta_x \langle H^{(1)} \rangle}{\partial_\lambda \langle H^{(1)} \rangle} = \frac{\delta \lambda}{\delta x} \cdot \frac{\mathrm{Tr}\left[\left(\rho_{x+\delta x} - \rho_{x-\delta x}\right)\cdot H^{(1)} \right]}{\mathrm{Tr}\left[\left(\rho_{\lambda+\delta \lambda} - \rho_{\lambda-\delta \lambda}\right)\cdot H^{(1)} \right]}.
\end{equation}
For the special case $x = \lambda$, the ratio trivially reduces to unity, recovering the standard QCRB.

\subsubsection{3.~Identifying the tight bound for the two-qubit model}
Based on the numerical results for QFI $\mathcal{F}_x(t)$ and response functions $\Delta_x\langle H^{(1)}\rangle(t)$, we now compute the observable-dependent precision bounds for the target parameter $\theta=\lambda$. According to Eq. (4) of the main text, the tightest precision bound for estimating $\lambda$ using the local observable $H^{(1)}$ is determined via
\begin{equation}\label{eq:bound_lambda}
    \mathrm{Var}(\lambda)~\ge~\max_{x \in \{t,\omega, \lambda, \gamma\}}\left\{\frac{1}{\mathcal{F}_x}\left|\frac{\Delta_x\langle H^{(1)}\rangle}{\partial_{\lambda}\langle H^{(1)}\rangle}\right|^2\right\}~=~\max_{x \in \{t,\omega, \lambda, \gamma\}}\left\{\mathcal{B}_x\right\}.
\end{equation}
For notational convenience, we denote $\mathcal{B}_x\equiv \mathcal{F}_x^{-1}\left|\Delta_x\langle H^{(1)}\rangle/\partial_{\lambda}\langle H^{(1)}\rangle\right|^2$. The bound corresponding to the target parameter $\lambda$ (dashed curve) is just the standard QCRB $\mathcal{B}_{\lambda}=\mathcal{F}_\lambda^{-1}$.

\begin{figure}[h]
 \centering
\includegraphics[width=0.75\columnwidth]{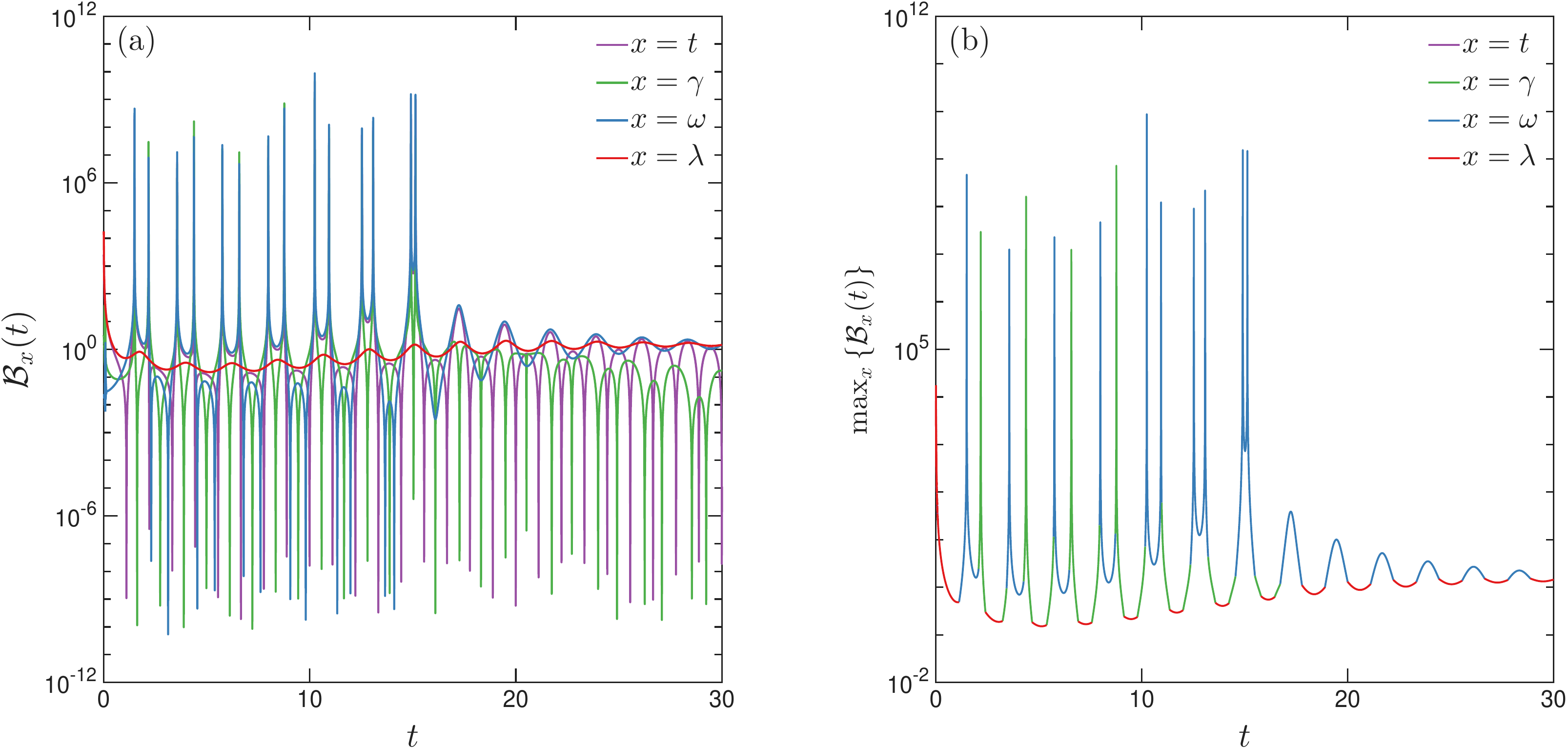} 
 \caption{(a) Dynamics of $\mathcal{B}_x = \mathcal{F}_x^{-1}|\Delta_x\langle H^{(1)}\rangle/\partial_\lambda\langle H^{(1)}\rangle|^2$ for different parameters $x=t$ (purple), $\gamma$ (green), $\omega$ (blue), and $\lambda$ (orange). (b) Time evolution of the tightest bound $\max_x\{\mathcal{B}_x(t)\}$ (colored). The initial state is $\rho(0)=\ket{gg}\bra{gg}$. Parameters are $\omega=0.5$, $\lambda=1$, $\gamma=0.2$, $\delta \omega =\delta \lambda =\delta \gamma = 0.001$ and $\delta t = 0.0001$.}
\protect\label{fig:Bounds}
\end{figure}

The time-dependent behaviors of the candidate bounds $\{\mathcal{B}_x\}$ and the resulting tightest bound are shown in Fig.~\ref{fig:Bounds}, illustrating the intricate competition among different bounds in constraining the estimation precision.
Fig.~\ref{fig:Bounds}~(a) displays the time evolution of all candidate bounds $\{\mathcal{B}_x(t)\}$ with $x\in \{t,\omega, \lambda, \gamma\}$. Interestingly, all bounds exhibit oscillatory behavior, yet with markedly different amplitudes. From Fig.~\ref{fig:Bounds}~(a), one already observes that the QCRB $\mathcal{B}_{\lambda}$ is not always the tightest bound, especially at short times. 

To identify the tight bound throughout the dynamics, we extract the maximum over $x$ at each time point according to Eq. (\ref{eq:bound_lambda}). The result is presented in Fig.~\ref{fig:Bounds} (b) which is also the tight bound showed in Fig. 4 of the main text. At short times, the tight bound alternates among $\mathcal{B}_{\lambda}$ (QCRB), $\mathcal{B}_{\omega}$ and $\mathcal{B}_{\gamma}$. For later times $t>16$, the competition narrows to an alternation between $\mathcal{B}_{\lambda}$ (QCRB) and $\mathcal{B}_{\omega}$. Consequently, depending on the measurement time, the ultimate precision limit can differ substantially from the standard QCRB. 

\section{III.~Spontaneous emission model}
As an additional example, we analyze an amplitude-damping model that describes the spontaneous emission of a two-level atom~\cite{Manzano.20.AP}. The evolution of the density matrix of this open two-level atom is governed by the following quantum Lindblad master equation
\begin{equation}\label{eq:lindblad_add}
    \partial_t \rho = - i \left[
    \frac{\omega}{2}\sigma_z, \rho \right] + \Gamma \left(\sigma_{-} \rho \sigma_{+} - \frac{1}{2} \{\sigma_{+} \sigma_{-}  , \rho \}\right).
\end{equation}
Here, $\omega$ denotes the energy splitting of the two‑level atom, and $\Gamma$ is the spontaneous emission rate. Hence, the complete parameter set for this model is $\{t,\omega,\Gamma\}$. Using the parametrization in Eq. (\ref{eq:rho_expression}), the solution of Eq. (\ref{eq:lindblad_add}) can be obtained as
\begin{equation}\label{eq:solution_add}
    \begin{aligned}
    \frac{1}{2}+p_t~=~&  \left(\frac{1}{2}+p_0\right)e^{-\Gamma t}, \\
    f_t~=~&f_0 e^{-i\omega t} e^{-\frac{1}{2}\Gamma t},
    \end{aligned}
\end{equation}
where $p_0$ and $f_0$ denote the initial values at $t=0$. 

\subsubsection{1.~Sum rule for the QFI}
In the eigenbasis of $\rho$, the symmetric logarithmic derivative operators for the parameters $\omega$, $\Gamma$ and $t$ take the form
\begin{equation}
    U^\dagger L_\omega U~=~a \sigma_y,~~U^\dagger L_\Gamma U~=~b\mathbb{I}+c\sigma_x+d\sigma_z,~~U^\dagger L_t U~=~b'\mathbb{I}+c'\sigma_x+a' \sigma_y+d'\sigma_z,
\end{equation}
where the time‑dependent coefficients are
\begin{equation}\label{eq:s61}
    a~=~-2t|f_t|,~~
    b~=~t\frac{p_t(\frac{1}{2}+p_t)+\frac{1}{2}|f_t|^2}{\mu_+\mu_-},~~
    c~=~-t\frac{|f_t|(1+p_t)}{\sqrt{p_t^2+|f_t|^2}},~~
    d~=~-t\frac{p_t(\frac{1}{2}+p_t)+\frac{1}{2}|f_t|^2}{2\mu_+\mu_-\sqrt{p_t^2+|f_t|^2}},
\end{equation}
and $a' = (a\omega) /t$, $b' = (b\Gamma)/t$, $c' = (c\Gamma)/t$, $d' = (d\Gamma)/t$. In above expressions, $\mu_\pm = 1/2 - \sqrt{p_t^2+|f_t|^2}$ [see Eq.~(\ref{eq:eigenvalue_rho})] are eigenvalues of the density matrix $\rho$, which are inherently positive. Using Eq. (\ref{eq:Fx_simple}) we obtain the diagonal QFI
\begin{equation}
    \mathcal{F}_\omega~=~a^2,~~\mathcal{F}_\Gamma~=~c^2+d^2-b^2,~~\mathcal{F}_t~=~a'^2+c'^2+d'^2-b'^2,
\end{equation}
while the off‑diagonal elements vanish, $\mathcal{F}_{\omega \Gamma}=\mathcal{F}_{\Gamma\omega}=0$. Combining these expressions for the elements of the QFI matrix gives
\begin{equation}
    t^2\mathcal{F}_t~=~\omega^2 \mathcal{F}_\omega +\Gamma^2\mathcal{F}_\Gamma,
\end{equation}
which confirms the validity of Eq. (3) in the main text for this dissipative model.

\subsubsection{2.~Tightest bound for observable \texorpdfstring{$\sigma_z$}{sigma z}}
We now determine the tightest bound for a given observable according to Eq. (4) of the main text, beginning with $\mathcal{O} = \sigma_z$. From the solution in Eq. (\ref{eq:solution_add}), we have $\langle \sigma_z\rangle = 2p_t$, whose derivatives are
\begin{equation}
    \partial_\omega \langle \mathcal\sigma_z\rangle~=~ 0,~~\partial_\Gamma \langle \mathcal\sigma_z\rangle~=~-2t\left( \frac{1}{2}+p_t\right),~~\partial_t \langle \mathcal\sigma_z\rangle~=~-2\Gamma\left( \frac{1}{2}+p_t\right).
\end{equation}
Applying Eq. (4) of the main text to the parameter-observable pair $(\theta,\sigma_z)$, the tightest bound is determined by
\begin{equation}
    \begin{aligned}
    \max_{x \in \{t, \omega, \Gamma\}} \left
    \{\frac{1}{\mathcal{F}_x}\left| \frac{\partial_x \langle \sigma_z\rangle}{\partial_\theta \langle \sigma_z\rangle}\right|^2\right\}~=~ &\frac{1}{\left| \partial_\theta \langle \sigma_z\rangle\right|^2}\max \left
    \{\frac{\left| \partial_\omega \langle \sigma_z\rangle\right|^2}{\mathcal{F}_\omega},~~\frac{\left| \partial_\Gamma \langle \sigma_z\rangle\right|^2}{\mathcal{F}_\Gamma},~~ \frac{\left| \partial_t \langle \sigma_z\rangle\right|^2}{\mathcal{F}_t}\right\}\\
    ~=~&\frac{4\left( \frac{1}{2}+p_t\right)^2}{|\partial_\theta \langle \sigma_z\rangle|^2} \max\left\{0,~~
    \frac{t^2}{c^2+d^2-b^2},~~\frac{\Gamma^2}{a'^2+c'^2+d'^2-b'^2}\right\}.
    \end{aligned}
\end{equation}
Noting that 
\begin{equation}
    \frac{\Gamma^2}{a'^2+c'^2+d'^2-b'^2}~=~\frac{t^2}{\left(\frac{\omega}{\Gamma}a\right)^2+c^2+d^2-b^2} \leq \frac{t^2}{c^2+d^2-b^2},
\end{equation}
we immediately obtain the tightest bound
\begin{equation}
    \max_{x \in \{\omega, \Gamma, t\}} \left
    \{\frac{1}{\mathcal{F}_x}\left| \frac{\partial_x \langle \sigma_z\rangle}{\partial_\theta \langle \sigma_z\rangle}\right|^2\right\}~=~\frac{1}{\mathcal{F}_\Gamma}\left| \frac{\partial_\Gamma \langle \sigma_z\rangle}{\partial_\theta \langle \sigma_z\rangle}\right|^2.
\end{equation}

\subsubsection{3.~Tightest bound for observable \texorpdfstring{$\sigma_+$}{sigma +}}
For the observable $\mathcal{O} = \sigma_+$, we have $\langle \sigma_+\rangle = f_t^*$, with derivatives
\begin{equation}\label{eq:Model_II_sigma_p}
    \partial_\omega \langle \mathcal\sigma_+\rangle~=~ (it)f_t^*,~~\partial_\Gamma \langle \mathcal\sigma_+\rangle~=~-\frac{t}{2}f_t^*,~~\partial_t \langle \mathcal\sigma_+\rangle~=~\left(-\frac{\Gamma}{2}+i\omega\right)f_t^*.
\end{equation}
Applying Eq. (4) of the main text to the parameter-observable pair $(\theta,\sigma_+)$, the tightest bound is determined by
\begin{equation}\label{eq:s69}
    \begin{aligned}
    \max_{x \in \{t,\omega, \Gamma\}} \left
    \{\frac{1}{\mathcal{F}_x}\left| \frac{\partial_x \langle \sigma_+\rangle}{\partial_\theta \langle \sigma_+\rangle}\right|^2\right\}~=~ &\frac{1}{\left| \partial_\theta \langle \sigma_+\rangle\right|^2}\max \left
    \{\frac{\left| \partial_\omega \langle \sigma_+\rangle\right|^2}{\mathcal{F}_\omega},~~\frac{\left| \partial_\Gamma \langle \sigma_+\rangle\right|^2}{\mathcal{F}_\Gamma},~~ \frac{\left| \partial_t \langle \sigma_+\rangle\right|^2}{\mathcal{F}_t}\right\}\\
    ~=~&\frac{|f_t|^2}{|\partial_\theta \langle \sigma_+\rangle|^2} \max\left\{\frac{t^2}{a^2},~~
    \frac{t^2/4}{c^2+d^2-b^2},~~
    \frac{\omega^2+(\Gamma^2/4)}{a'^2+c'^2+d'^2-b'^2}\right\} \\
    ~=~&\frac{t^2|f_t|^2}{4|\partial_\theta \langle \sigma_+\rangle|^2} \max\left\{\frac{4}{a^2},~~
    \frac{1}{c^2+d^2-b^2},~~
    \frac{4\omega^2+\Gamma^2}{\omega^2a^2+\Gamma^2(c^2+d^2-b^2)}\right\}.
    \end{aligned}
\end{equation}
Observing that
\begin{equation}
    \max\left\{\frac{4}{a^2},~~
    \frac{1}{c^2+d^2-b^2}\right\} \geq 
    \frac{4\omega^2+\Gamma^2}{\omega^2a^2+\Gamma^2(c^2+d^2-b^2)}\geq \min\left\{\frac{4}{a^2},~~\frac{1}{c^2+d^2-b^2}\right\},
\end{equation}
Eq. (\ref{eq:s69}) simplifies to
\begin{equation}\label{eq:s71}
    \max_{x \in \{t, \omega, \Gamma\}} \left
    \{\frac{1}{\mathcal{F}_x}\left| \frac{\partial_x \langle \sigma_+\rangle}{\partial_\theta \langle \sigma_+\rangle}\right|^2\right\}
    ~=~\frac{t^2|f_t|^2}{4|\partial_\theta \langle \sigma_+\rangle|^2} \max\left\{\frac{4}{a^2},~~
    \frac{1}{c^2+d^2-b^2}\right\}.
\end{equation}
Thus we only need to compare the two quantities $4/a^2$ and $1/(c^2+d^2-b^2)$, i.e., to determine the sign of $\Delta \equiv c^2+d^2-b^2-(a^2/4)$. Substituting the explicit forms of the coefficients in Eq. (\ref{eq:s61}) yields, after algebra,
\begin{equation}\label{eq:s72}
    \begin{aligned}
    \Delta 
    ~=~&t^2 \frac{\left[p_t \left( \frac{1}{2}+p_t\right)+\frac{1}{2}|f_t|^2 \right]^2}{\mu_+\mu_-(p_t^2+|f_t|^2)}+ t^2 \frac{|f_t|^2(1+p_t)^2}{p_t^2+|f_t|^2}-t^2|f_t|^2 \\
    ~\ge~&t^2 \frac{\left[p_t \left( \frac{1}{2}+p_t\right)-\frac{1}{2}|f_t|^2 \right]^2}{\mu_+\mu_-(p_t^2+|f_t|^2)}+ t^2 \frac{|f_t|^2(1+p_t)^2}{p_t^2+|f_t|^2}-t^2|f_t|^2~\equiv~\Delta' \\
    ~=~&\frac{t^2}{\mu_+\mu_-(p_t^2+|f_t|^2)} \left\{\left[p_t \left( \frac{1}{2}+p_t\right)-\frac{1}{2}|f_t|^2 \right]^2+ \mu_+\mu_-|f_t|^2\left[ 2\left( \frac{1}{2}+p_t\right)-|f_t|^2\right] \right\}.
    \end{aligned}
\end{equation}
Here the inequality follows from $(A+B)^2\ge (A-B)^2$ for non-negative $A=p_t(\frac{1}{2}+p_t)$ and $B=\frac{1}{2}|f_t|^2$. Consequently, to establish $\Delta\ge 0$, it suffices to prove $\Delta'\ge 0$. To analyze the sign of $\Delta'$ systematically, we introduce the abbreviations $A = \exp(-\Gamma t)$, $P = 1/2+p_0$ and $F = |f_0|^2$ and define $z=1-F/P$ (clearly, $0\leq z \leq 1$) and $y=PA$. Using the solution Eq. (\ref{eq:solution2}), Eq. (\ref{eq:s72}) can be recast as  
\begin{equation}\label{eq:s73}
    \frac{\mu_+\mu_-(p_t^2+|f_t|^2)}{t^2 F^2A^2 x^2}\Delta'~=~z^2 y^2 +(-z^3+2z-2)y + \frac{(2-z)^2}{4},
\end{equation}
where the positivity of the density matrix implies $F/P \leq 1-PA = 1-y$, hence $0\leq y \leq z$. Define the bivariate function
\begin{equation}
    \mathcal{G}(y,z)~=~z^2 y^2 +(-z^3+2z-2)y + \frac{(2-z)^2}{4},
\end{equation}
whose sign, via Eq. (\ref{eq:s73}), determines the sign of the quantity $\Delta'$. At the boundary of $y$, we have $\mathcal{G}(0,z)= (2-z)^2/4 \geq 0$ and $\mathcal{G}(z,z)= (3z-2)^2/4 \geq 0$. For interior points $0<y<z$ we treat $\mathcal{G}(y,z)$ as a quadratic function in $y$. Its quadratic coefficient $z^{2}\geq 0$ makes the parabola convex (or linear when $z=0$). The symmetry axis $Y_{\mathrm{sym}}$ and discriminant $\mathcal{Y}$ of this quadratic function are  
\begin{equation}
    Y_{\mathrm{sym}}~=~\frac{z^3-2z+2}{2z^2},~~
    \mathcal{Y}~=~(z-1)(z+1)\left[z^4 - 4(z-1)^2\right].
\end{equation}
The condition that the symmetry axis lies within $0 < Y_{\mathrm{sym}} < z$ reduces to
\begin{equation}\label{eq:s76}
    z^3+2z-2~>~0.
\end{equation}
We note that the cubic function $f(z)=z^3+2z-2$ has a single real zero point $z_0$; by Cardano’s formula it is
\begin{equation}
    z_0~=~\sqrt[3]{1+\sqrt{\frac{35}{27}}} + \sqrt[3]{1-\sqrt{\frac{35}{27}}}~>~0.771.
\end{equation}
Thus inequality Eq. (\ref{eq:s76}) holds for $z_0\leq z\leq1$. At the symmetry axis $y=Y_{\mathrm{sym}}$ the function attains its minimum $\mathcal{G}(Y_{\mathrm{sym}},z)=-\mathcal{Y}/(4z^{2})$. The discriminant $\mathcal{Y}$ vanishes at the four real zeros $z_1=-1$, $z_2=-1-\sqrt{3}$, $z_3=-1+\sqrt{3}$, $z_4 = 1$, with $z_0>z_3$. In the interval $(z_0,1)$ the quartic factor $\mathcal{Q}(z)=z^{4}-4(z-1)^{2}$ is strictly increasing, therefore
\begin{equation}
    \mathcal{Q}(z)~\geq~\mathcal{Q}(z_{0})~>~
    \mathcal{Q}(z_{3})~=~0.
\end{equation}
implying $\mathcal{Y}(z)>0$ and consequently $\mathcal{G}(Y_{\mathrm{sym}},z)<0$ for $z_0< z<1$. However, since $\mathcal{G}(0,z)\geq 0$ and $\mathcal{G}(z,z)\geq 0$, convexity guarantees $\mathcal{G}(y,z)\ge 0$ for all $0\leq y\leq z$ and $0\leq z\leq 1$. Therefore, we obtain
\begin{equation}
    \mathcal{G}(y,z)~\geq~0,
\end{equation}
and from Eq. (\ref{eq:s71}) we finally obtain 
\begin{equation}
    \max_{x\in\{\omega,\Gamma,t\}} 
    \left\{ \frac{1}{\mathcal{F}_x} \left| \frac{\partial_x \langle \sigma_+ \rangle}{\partial_\theta \langle \sigma_+ \rangle} \right|^2 \right\} 
    = \frac{1}{\mathcal{F}_\omega} \left| \frac{\partial_\omega \langle \sigma_+ \rangle}{\partial_\theta \langle \sigma_+ \rangle} \right|^2.
\end{equation}

\subsubsection{4.~Tightest bound for observable \texorpdfstring{$\sigma_-$}{sigma -}}
For the observable $\mathcal{O} = \sigma_-$ is chosen, we have $\langle \sigma_-\rangle = f_t$, with derivatives
\begin{equation}
    \partial_\omega \langle \mathcal\sigma_-\rangle~=~ (-it)f_t,~~\partial_\Gamma \langle \mathcal\sigma_-\rangle~=~-\frac{t}{2}f_t,~~\partial_t \langle \mathcal\sigma_-\rangle~=~\left(-\frac{\Gamma}{2}-i\omega\right)f_t.
\end{equation}
Because $|\partial_x \langle \mathcal\sigma_-\rangle|^2=|\partial_x \langle \mathcal\sigma_+\rangle|^2$ for each $x \in \{t,\omega, \Gamma\}$, the analysis follows identically to the $\sigma_+$ case, yielding the analogous conclusion,
\begin{equation}
    \max_{x\in\{t,\omega,\Gamma\}} 
    \left\{ \frac{1}{\mathcal{F}_x} \left| \frac{\partial_x \langle \sigma_- \rangle}{\partial_\theta \langle \sigma_- \rangle} \right|^2 \right\} 
    = \frac{1}{\mathcal{F}_\omega} \left| \frac{\partial_\omega \langle \sigma_- \rangle}{\partial_\theta \langle \sigma_- \rangle} \right|^2.
\end{equation}

To summarize, after evaluating all candidate bounds in Eq.~(4) of the main text for the accessible observables $\{\sigma_z, \sigma_{\pm}\}$, we identify the tightest bound for each observable-parameter pair $(\mathcal{O}, \theta)$ under arbitrary initial states. The results are summarized in Table~\ref{tab:damping}. We observe that in some cases the QCRB remains tight even when suboptimal observables are used, highlighting that our bound is intrinsically a quantum bound despite the fixed observable. In other cases, however, the tight bound deviates from the corresponding QCRB. Thus, the form of the tight bound varies between different observables.

\begin{table}[t!]
\renewcommand{\arraystretch}{2.5}
\centering
\caption{Tightest bound [Eq. (4) in the main text] for estimating a parameter $\theta\in\{t,\omega,\Gamma\}$ in the spontaneous emission model under a given observable $\mathcal{O}$.}
\begin{tabular}{c|c|c|c}
\hline\hline
\hspace*{0.5cm}{}\hspace*{0.5cm} & \hspace*{0.5cm}$\mathcal{O}=\sigma_z$\hspace*{0.5cm} & \hspace*{0.5cm}$\mathcal{O}=\sigma_-$\hspace*{0.5cm} & \hspace*{0.5cm}$\mathcal{O}=\sigma_+$\hspace*{0.5cm}  \\ 
\hline
\hspace*{0.5cm}$\theta = t$\hspace*{0.5cm} & $\frac{1}{\mathcal{F}_{\Gamma}}\left| \frac{\partial_{\Gamma} \langle \sigma_z\rangle}{\partial_{t} \langle \sigma_z\rangle}\right|^2$ & $\frac{1}{\mathcal{F}_\omega}\left| \frac{\partial_\omega \langle \sigma_-\rangle}{\partial_{t} \langle \sigma_-\rangle}\right|^2$ & $\frac{1}{\mathcal{F}_\omega}\left| \frac{\partial_\omega \langle \sigma_+\rangle}{\partial_{t} \langle \sigma_+\rangle}\right|^2$  \\ 
\hline
\hspace*{0.5cm}$\theta = \omega$\hspace*{0.5cm} & $\frac{1}{\mathcal{F}_{\Gamma}}\left| \frac{\partial_{\Gamma} \langle \sigma_z\rangle}{\partial_{\omega} \langle \sigma_z\rangle}\right|^2$ & $\frac{1}{\mathcal{F}_\omega}$ & $\frac{1}{\mathcal{F}_\omega}$  \\ 
\hline
\hspace*{0.5cm}$\theta = \Gamma$\hspace*{0.5cm} & $\frac{1}{\mathcal{F}_{\Gamma}}$ & $\frac{1}{\mathcal{F}_\omega}\left| \frac{\partial_\omega \langle \sigma_-\rangle}{\partial_{\Gamma} \langle \sigma_-\rangle}\right|^2$ & $\frac{1}{\mathcal{F}_\omega}\left| \frac{\partial_\omega \langle \sigma_+\rangle}{\partial_{\Gamma} \langle \sigma_+\rangle}\right|^2$ \\ 
\hline\hline
\end{tabular}
\label{tab:damping}
\end{table}

\end{document}